\documentclass[sigconf,nonacm,balance=false]{acmart}
\usepackage{amsmath}
\usepackage{booktabs}
\usepackage{microtype}
\usepackage[checkfootnote,checkfloat,nospread]{flushend}
\setcopyright{none}
\title[CRIM: Composable and Reusable Item Memory]{CRIM: Composable and Reusable Item Memory for Efficient Personalized Retrieval}

\author{Haolin Li}
\authornote{Haolin Li and Haipeng Zhang contributed equally to this work.}
\affiliation{\institution{Tsinghua University}\city{Beijing}\country{China}}
\affiliation{\institution{Alibaba International Digital Commerce Group}\city{Beijing}\country{China}}
\email{hl-li23@mails.tsinghua.edu.cn}

\author{Haipeng Zhang}
\authornotemark[1]
\affiliation{\institution{Alibaba International Digital Commerce Group}\city{Beijing}\country{China}}

\author{Mang Li}
\affiliation{\institution{Alibaba International Digital Commerce Group}\city{Beijing}\country{China}}

\author{Zhaohui Xia}
\affiliation{\institution{Alibaba International Digital Commerce Group}\city{Beijing}\country{China}}

\author{Xun Zhu}
\affiliation{\institution{Tsinghua University}\city{Beijing}\country{China}}

\author{Miao Li}
\affiliation{\institution{Tsinghua University}\city{Beijing}\country{China}}

\author{Yu Zhang}
\authornote{Corresponding authors: Yu Zhang and Ji Wu.}
\affiliation{\institution{Alibaba International Digital Commerce Group}\city{Beijing}\country{China}}

\author{Ji Wu}
\authornotemark[2]
\affiliation{\institution{Tsinghua University}\city{Beijing}\country{China}}
\renewcommand{\shortauthors}{Li et al.}

\begin{document}
\begin{abstract}
Large language models incur substantial training and inference costs in personalized search and recommendation, as growing interaction histories lengthen request inputs and repeatedly expose the same item text to the model. We propose \textbf{C}omposable and \textbf{R}eusable \textbf{I}tem \textbf{M}emory (CRIM), a framework that separates item-level semantic encoding from request-level history modeling. A small set of learnable summary tokens compresses each item into compact continuous memories, while segment-level readout tokens (REC tokens) introduce predictive supervision within the interaction history. These memories are independent of users and queries and can therefore be reused across requests. At inference time, CRIM composes cached memories in interaction order and incorporates an optional query to produce a personalized retrieval representation, reducing repeated item-text encoding while retaining request-dependent contextual modeling. We further introduce a joint training mechanism that combines cached memories with sampled live re-encoding. Re-encoding only a few history items per request allows the retrieval objective to jointly optimize the item and request encoders at limited computational cost, while periodic cache refreshes maintain consistency. Extensive evaluation and cost analysis on MovieLens-1M, Qilin, and our new LazSeq dataset show that CRIM achieves state-of-the-art retrieval performance with substantially lower inference costs at the same backbone scale, while limiting item re-encoding during joint training. We will publicly release our code and dataset.
\end{abstract}

\maketitle
\hypersetup{pdfauthor={Haolin Li, Haipeng Zhang, Mang Li, Zhaohui Xia, Xun Zhu, Miao Li, Yu Zhang, Ji Wu}}
\raggedbottom
\begin{figure}[!t]
  \centering
  \includegraphics[width=\columnwidth]{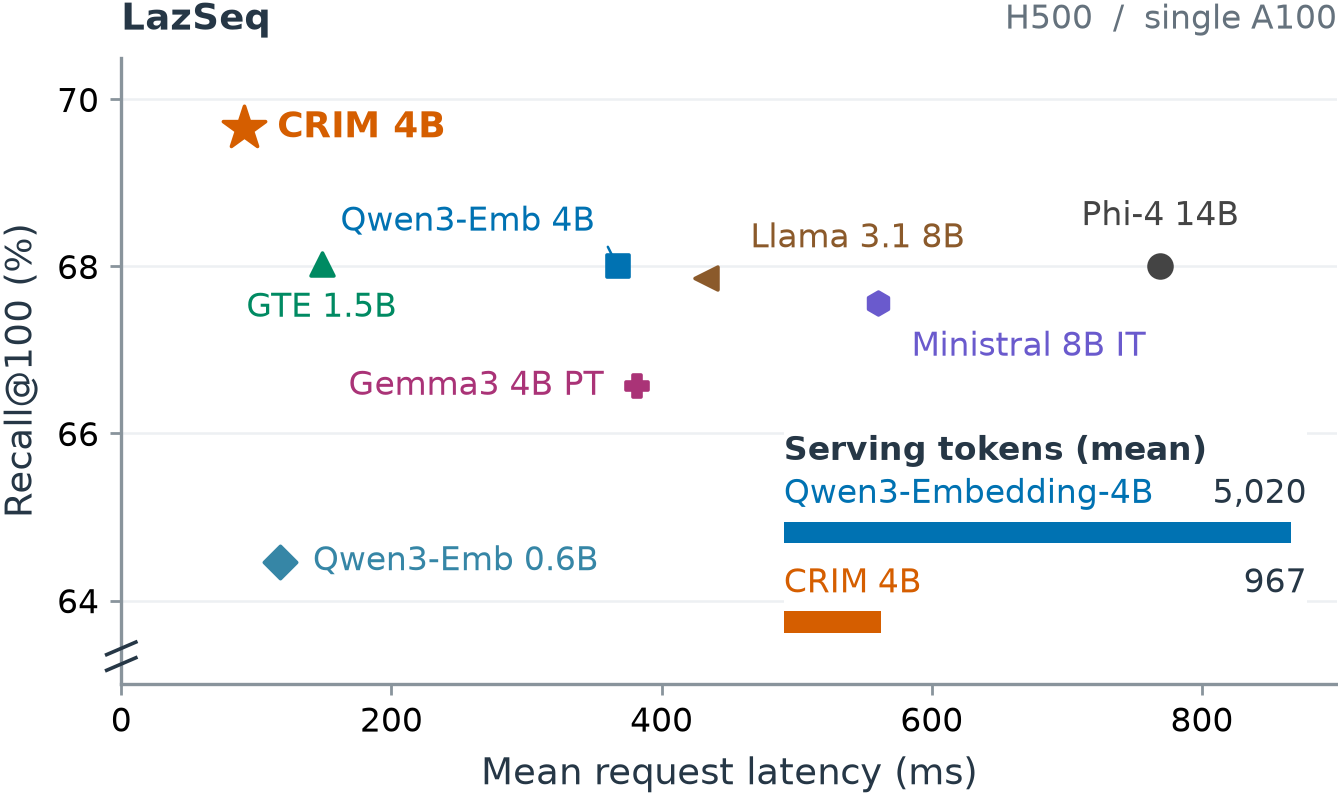}
  \caption{Recall@100 versus mean request latency on LazSeq (single A100, histories up to 500 interactions). The inset reports mean serving input lengths, including memory slots, REC tokens, queries, and special tokens.}
  \Description{A scatter plot compares eight models on LazSeq. CRIM has the highest Recall at 100 and the lowest mean request latency among the displayed models. The inset shows 967 mean serving input positions for CRIM and 5,020 for Qwen3-Embedding-4B.}
  \label{fig:quality-efficiency}
\end{figure}

\section{Introduction}
\label{sec:introduction}

Personalized Web search and recommendation retrieve relevant items from large catalogs using a user's ordered interaction history and, when available, a current query. Large language models provide rich semantic representations of items and queries, but growing histories increase both training and inference costs. Repeatedly encoding the same items across users and requests adds further overhead. Efficient personalized retrieval therefore requires addressing three challenges: reusing item semantics while retaining request-dependent contextual modeling; preserving both semantic information and interaction patterns in compressed representations; and jointly optimizing item and retrieval encoders when cached representations reduce the amount of differentiable computation.

Existing methods address different aspects of these challenges. Recformer~\cite{recformer} jointly models item text and interaction sequences, while processing history text for each request. Token Factory~\cite{tokenfactory} converts heterogeneous features into soft tokens and compresses recommendation inputs, and ICAE~\cite{icae} learns compact memories through autoencoding and language modeling. These approaches demonstrate the value of compression. For personalized retrieval, however, an item representation must also remain useful when composed into different histories, and text reconstruction alone does not directly supervise interaction patterns. Representation caching offers a complementary way to reduce repeated computation. GNNAutoScale~\cite{gnnautoscale}, for example, reuses historical embeddings in graph neural networks. Applying caching to language-model-based item encoding requires coordinating encoder updates with the gradient paths available through cached representations, without allowing full-history re-encoding to consume the savings from reuse.

Our central idea is to separate stable item semantics from dynamic request context. We propose \textbf{C}omposable and \textbf{R}eusable \textbf{I}tem \textbf{M}emory (CRIM). An item memory encoder (IME) independently maps each item to a compact block of continuous memories shared across users and requests. A request encoder (RE) composes these blocks in interaction order and incorporates an optional query to generate a personalized retrieval representation. This design moves repeated item-text encoding offline while preserving contextual modeling of the current history and information need. Preserving item-text semantics allows retrieval to use the LLM's pretrained semantic knowledge alongside learned interaction patterns.

To support semantic compression and sequence learning together, CRIM combines learnable summary tokens with segment-level readout tokens, termed REC tokens. Summary tokens encode each item's semantics into a small number of continuous vectors. REC tokens read out intermediate history prefixes to predict subsequent interactions, introducing supervision throughout the sequence. We further develop a joint training mechanism that combines cached memories with sampled live re-encoding. Only a few history positions are re-encoded for each request, allowing retrieval gradients to update the IME through those positions. The remaining positions reuse cached memories and participate in RE training. Periodic refreshes keep the memory cache aligned with the evolving encoder. This mechanism allows item memories to adapt to the retrieval objective while limiting the cost of re-encoding long histories.

We introduce LazSeq, a benchmark for query-conditioned retrieval with approximately 30,000 requests, a shared catalog of 49,993 distinct items, and histories of up to 500 interactions. Together with MovieLens-1M (ML1M)~\cite{movielens} and Qilin~\cite{qilin}, it supports an extensive evaluation across retrieval methods and task settings. CRIM achieves state-of-the-art retrieval performance. Compared with the Qwen3-Embedding-4B baseline~\cite{qwen3embedding}, it improves Recall@100 by 2.49\% on LazSeq and 33.61\% on ML1M, and NDCG@100 by 5.64\% on Qilin, all measured as relative improvements. As shown in Figure~\ref{fig:quality-efficiency}, CRIM combines strong retrieval quality with lower latency and fewer serving tokens on LazSeq at the same backbone scale. Our cost analysis relates offline investment and request computation to the volume of requests that reuse item memories.

Our contributions are fourfold:
\begin{itemize}
  \item \textbf{Composable and reusable item memory.} We propose a novel framework that separates item-level semantic encoding from request-level history modeling, enabling reuse across requests while retaining personalized contextual modeling.
  \item \textbf{Semantic compression with sequence supervision.} We combine summary tokens for compact item representations with REC tokens that introduce predictive supervision within interaction histories.
  \item \textbf{Efficient joint training.} We combine cached memories with sampled live re-encoding, allowing item memories to adapt to the retrieval objective while controlling re-encoding costs.
  \item \textbf{LazSeq and comprehensive evaluation.} We introduce a long-history retrieval dataset and evaluate retrieval quality and computational efficiency across three datasets, supported by a cost analysis of memory reuse.
\end{itemize}

\begin{figure*}[!t]
  \centering
  \includegraphics[width=\textwidth]{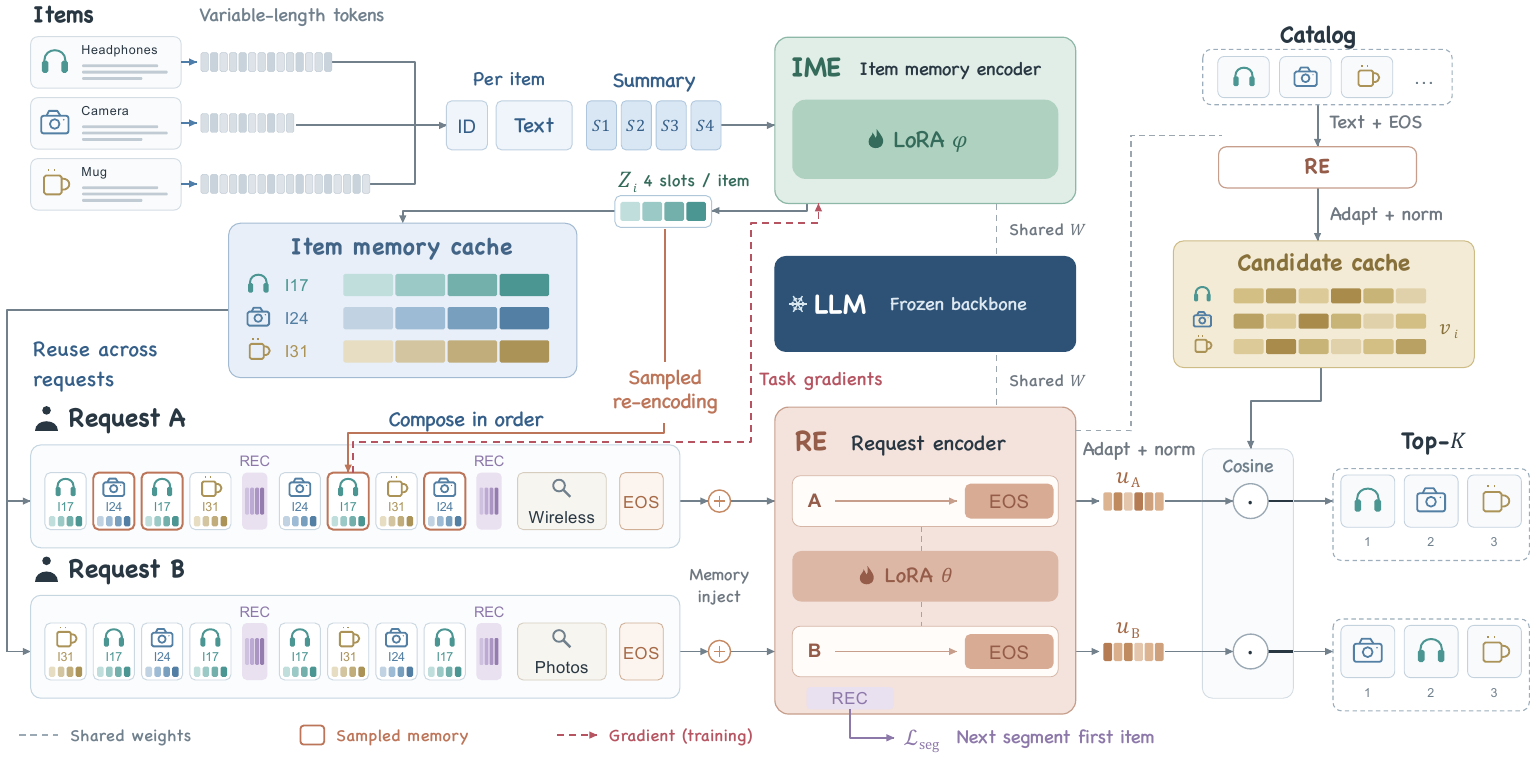}
  \caption{Overview of CRIM. The IME encodes each item into reusable continuous memories. The RE composes cached memory blocks in interaction order, incorporates an optional query, and produces a representation for retrieval against separately encoded catalog candidates.}
  \Description{The CRIM framework separates independent item memory encoding, reuse of cached memories in ordered histories, and request-conditioned retrieval.}
  \label{fig:overview}
\end{figure*}

\section{Related Work}
\label{sec:related}

\paragraph{Language models for personalized retrieval.}
Recformer~\cite{recformer} organizes item attributes as text sequences and jointly learns language representations and interaction patterns. TIGER~\cite{tiger} maps items to semantic identifiers and autoregressively predicts the next item. GEMS~\cite{gems} unifies search and recommendation through gradient subspace decomposition that coordinates task-specific optimization signals. Related adaptation methods include LiRA~\cite{lira2026}, which aligns low-resource languages in a shared semantic space for multilingual retrieval, and AgentDet~\cite{agentdet2026}, which coordinates detection agents through a shared blackboard and knowledge base for zero-/few-shot object detection. CRIM focuses on repeated item encoding in long histories, using shared item memories to support request-dependent composition and integrate history with an optional query in a dense retrieval framework.

\paragraph{Context compression and reusable item representations.}
Prefix-tuning~\cite{prefixtuning} and prompt tuning~\cite{prompttuning} learn continuous task prompts. Context compression instead encodes input content: LLMLingua~\cite{llmlingua} prunes prompt tokens, while continuous compressors retain information from long text in a small set of vectors. ICAE~\cite{icae} learns compact memory slots through autoencoding and language modeling, while EBMC~\cite{ebmc} pretrains a decoupled compressor using text reconstruction and completion. For industrial recommendation, Token Factory~\cite{tokenfactory} maps heterogeneous features to soft tokens and compresses behavioral inputs. Dual-purpose Semantic IDs~\cite{dualpurpose} combine discrete encoding with on-demand vector reconstruction to reduce item representation storage and transfer costs. CRIM uses individual items as the unit of reuse: summary tokens retain item semantics, and REC tokens introduce prediction objectives within interaction histories so that compact memories support composition and retrieval across users.

\paragraph{Efficient training with cached representations.}
Caching reduces repeated computation, but its effectiveness depends on gradient paths and representation refresh frequency. Graph Segment Training~\cite{gst} backpropagates through sampled graph segments while reusing historical embeddings for the remaining segments. REST~\cite{rest} reduces representation staleness by separating forward refreshes from parameter updates. CRIM applies partial differentiable computation and caching to language-model-based item memory. Each request re-encodes only a small sample of history positions, allowing the retrieval objective to jointly optimize the IME and RE; periodic refreshes maintain cache consistency. 

\section{Method}
\label{sec:method}

\subsection{Task Definition and Framework}
\label{sec:framework}
We study personalized retrieval from user interaction histories. Let $\mathcal I$ be an item catalog, where each item $i$ has a textual description $x_i$. A request consists of a chronologically ordered history $H=(i_1,\ldots,i_N)$ and an optional query $q$. The objective is to retrieve a relevant target item $y$ from the full catalog. Setting $q=\varnothing$ yields history-based recommendation without an explicit query.

The same item can appear in the histories of multiple users and requests. Its textual semantics can be reused, whereas user interests depend on the interaction order and current query. CRIM therefore separates item-level semantic encoding from request-level history modeling through two modules. The IME compresses an item's text and identity into a fixed number of continuous vectors. The RE composes these memories in interaction order and incorporates the optional query:
\begin{equation}
\begin{aligned}
Z_i &= C_\phi(x_i,e_i)\in\mathbb R^{K\times d},\\
u &= f_\theta(Z_{i_1},\ldots,Z_{i_N},q),\\
s(H,q,i) &= \frac{u^\top v_i}{\|u\|_2\|v_i\|_2}.
\end{aligned}
\label{eq:framework}
\end{equation}
Here, $e_i$ is an item identity prefix, $K$ is the number of memory vectors per item, $d$ is the hidden dimension, and $v_i$ is the candidate representation. Candidates are ranked by $s(H,q,i)$.

Item memories are encoded independently of users, histories, and queries, so they can be precomputed and cached. At inference time, CRIM retrieves the corresponding memory blocks in history order, retaining repeated interactions, and models the request context with the RE. Candidate vectors are constructed and cached separately for full-catalog retrieval. Figure~\ref{fig:overview} illustrates the framework.

\subsection{Item Memory Encoding and Semantic Pretraining}
\label{sec:pretraining}
To summarize item text, we append a shared set of $K$ learnable summary tokens with embeddings $S=(s_1,\ldots,s_K)$ to every item input. Let $E_i$ denote the embedded item identity prefix and $T(x_i)$ the embedded text template. Encoding yields
\begin{equation}
Z_i=\left[F_\phi\!\left(E_i\oplus T(x_i)\oplus S\right)\right]_{\mathrm{sum}},
\label{eq:memory}
\end{equation}
where $\oplus$ denotes concatenation and $F_\phi$ is a causal language model. The operator $[\cdot]_{\mathrm{sum}}$ extracts the summary-position hidden states after the final normalization layer. Causal attention allows these positions to read the preceding identity prefix and item text. All items share these $K$ learned embeddings, adding only $Kd$ summary-token parameters regardless of catalog size. The resulting memory vectors depend on item content.

We first perform semantic pretraining on the item catalog through reconstruction. A decoding sequence is appended after the summary tokens, with the item title as the reconstruction target. A restricted attention mask allows decoding positions to access only the summary positions and preceding decoding tokens, blocking direct access to the source text and identity prefix. Consequently, source information must pass through the summary tokens to support reconstruction.

Alongside reconstruction cross-entropy, we use the predictive distribution under full context as a distillation target. The same frozen language model produces this reference distribution with direct access to the source text. Let $\mathcal T$ contain the supervised decoding positions in a batch, $a_{i,t}$ be the target token, and $p^{\mathrm{mem}}_{i,t}$ and $p^{\mathrm{full}}_{i,t}$ be the distributions under restricted and full attention. The pretraining objective is
\begin{equation}
\begin{aligned}
\mathcal L_{\mathrm{pre}}
={}&\frac{1}{|\mathcal T|}\sum_{(i,t)\in\mathcal T}
\Bigl[-\log p^{\mathrm{mem}}_{i,t}(a_{i,t})\\
&+\lambda_{\mathrm{KL}}D_{\mathrm{KL}}\!\left(
\operatorname{sg}[p^{\mathrm{full}}_{i,t}]
\,\middle\|\,p^{\mathrm{mem}}_{i,t}\right)\Bigr],
\end{aligned}
\label{eq:stage-a}
\end{equation}
where $\operatorname{sg}$ stops gradients and $\lambda_{\mathrm{KL}}$ controls the distillation weight. Cross-entropy directly supervises title reconstruction; the KL term aligns predictions under the two context conditions.

\begin{figure}[!t]
\centering
  \includegraphics[width=0.94\columnwidth]{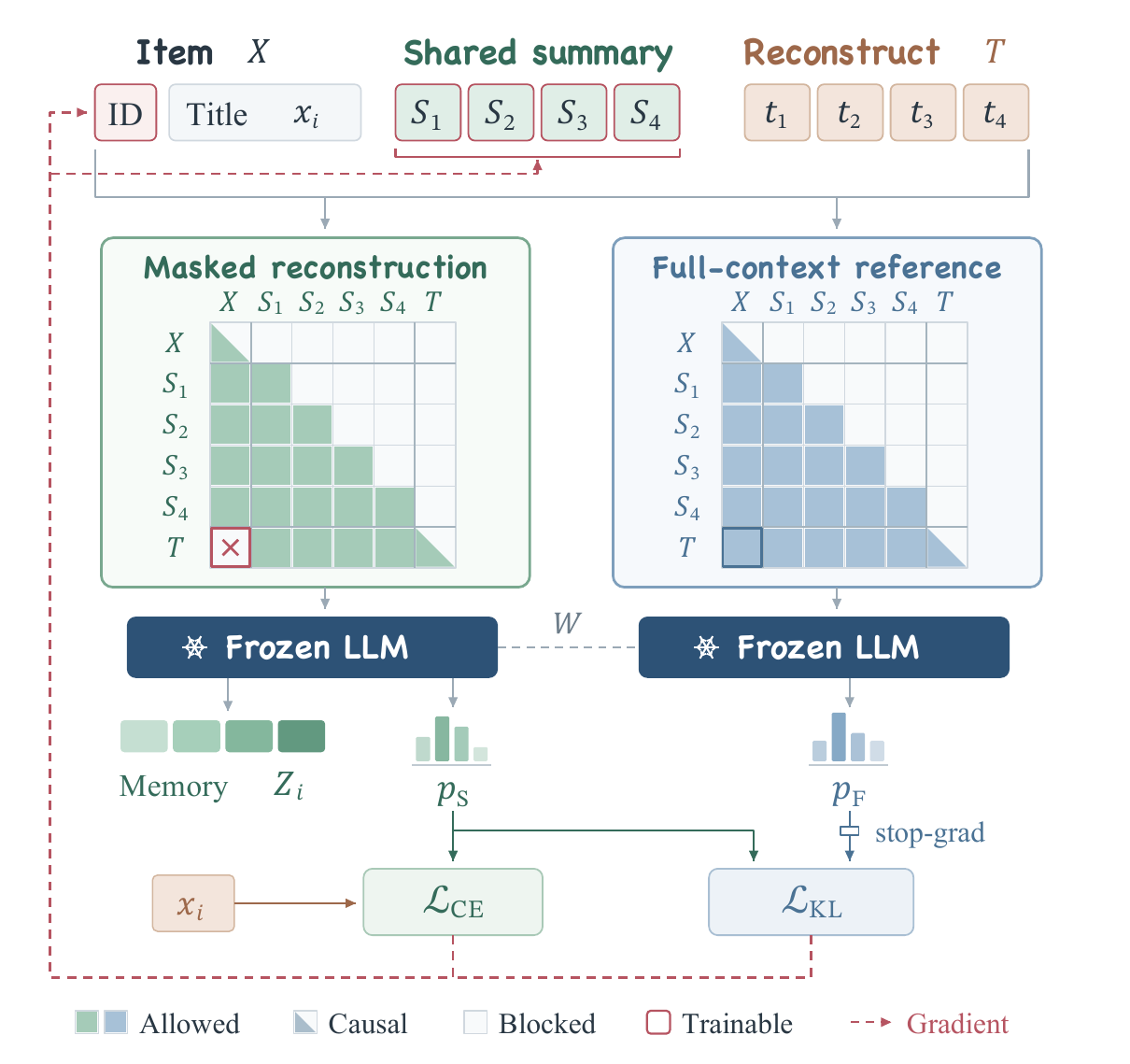}
  \caption{Semantic pretraining: summary tokens compress item identity and text under restricted-attention reconstruction and full-context distillation with a frozen backbone.}
  \Description{Summary tokens form a bottleneck between the source item and its reconstructed title. Reconstruction and distribution matching supervise the memory slots.}
  \label{fig:slot-learning}
\end{figure}

Semantic pretraining updates only summary-token embeddings, item identity embeddings, and their projection parameters. The language model backbone remains frozen, and no user interaction sequences are required. Figure~\ref{fig:slot-learning} illustrates this stage. These parameters provide a semantic initialization for the item memories, which are subsequently adapted to personalized retrieval through downstream supervision.

\subsection{Sequence-Aware Memory Composition}
\label{sec:composition}
Independent item memories retain item semantics, while personalized retrieval additionally requires modeling interaction order and evolving interests. The RE therefore composes memories chronologically and uses segment-level supervision to learn representations of different history prefixes.

The memory vectors enter the RE as continuous input embeddings through a learnable residual mapping. For the $k$-th memory vector $z_{i,k}$ of item $i$,
\begin{equation}
\bar z_{i,k}=z_{i,k}+W_{\mathrm{in}}z_{i,k}+a_k,
\label{eq:input}
\end{equation}
where $W_{\mathrm{in}}$ is a residual projection and $a_k$ is a learnable within-item memory-position embedding. Both are initialized to zero, so the RE initially receives the original memories and gradually learns an input representation suited to history modeling.

We divide the history into $J=\lceil N/b\rceil$ contiguous segments of at most $b$ items. A shared set of learnable segment readout tokens, termed REC tokens, is inserted after each segment. History memories use causal attention. The REC tokens of segment $j$ can attend to all memories up to that segment and earlier REC tokens within the same readout block. Subsequent history memories, query tokens, and the final retrieval position do not attend to REC hidden states. This provides additional supervision positions while preserving direct interactions among history memories.

For query-conditioned retrieval, query tokens follow the history and can attend to all history memories. The final EOS hidden state passes through a representation adapter and normalization to produce $u$. Inserting REC tokens does not advance the position indices of subsequent memories and query tokens, preserving their original positional relationships. The candidate branch reuses the RE's language model to encode item text followed by EOS; a candidate-side representation adapter and normalization produce $v_i$.

To supervise the history internally, each segment readout predicts the first item in the next segment. Let $r_j$ be obtained by averaging segment $j$'s REC hidden states, then applying a shared readout adapter and normalization. For a representation $h$, positive item $y$, and negative set $\mathcal N$, define
\begin{equation}
\ell(h,y;\mathcal N)=
-\log\frac{\exp(\operatorname{cos}(h,v_y)/\tau)}
{\exp(\operatorname{cos}(h,v_y)/\tau)+
\sum_{n\in\mathcal N}\exp(\operatorname{cos}(h,v_n)/\tau)},
\label{eq:contrastive}
\end{equation}
where $\tau$ is a temperature. The segment objective is
\begin{equation}
\mathcal L_{\mathrm{seg}}=
\frac{1}{J-1}\sum_{j=1}^{J-1}
\ell(r_j,i_{jb+1};\mathcal N_j).
\label{eq:segment}
\end{equation}
Here, $i_{jb+1}$ is the first item of the next segment. Negatives are sampled uniformly from the catalog, excluding the positive item identity, and candidate representations are cached. We set $\mathcal L_{\mathrm{seg}}=0$ when $J\leq1$. These intermediate prediction positions directly supervise the relationship between each history prefix and subsequent interactions.

\begin{figure}[!t]
\centering
  \includegraphics[width=0.80\columnwidth,trim=77.1bp 87.3bp 76.0bp 33.4bp,clip]{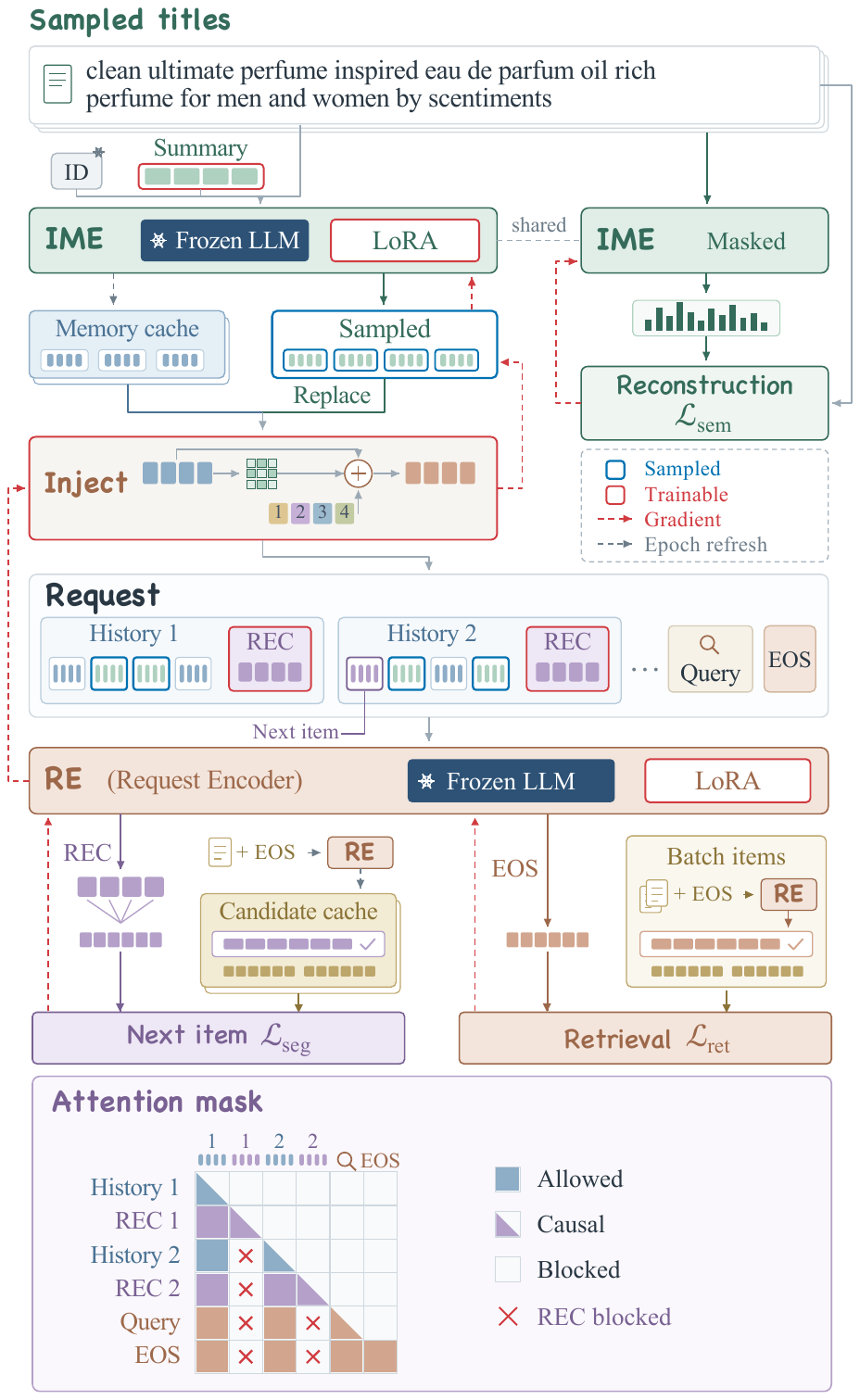}
  \caption{Joint training: sampled positions use live IME outputs; the others reuse cached memories. Retrieval, segment prediction, and reconstruction provide supervision.}
  \Description{The training pipeline combines a cached memory path and a sampled live IME path, preserving the full ordered history for the RE and refreshing cached representations after training epochs.}
  \label{fig:joint-training}
\end{figure}

\subsection{Joint Training with Cached and Sampled Memories}
\label{sec:joint}
Item memories must adapt to the retrieval objective, but live encoding of every history item makes training cost grow with history length. During joint training, we combine cached memories with sampled live re-encoding. The complete history participates in request modeling, while differentiable item encoding is limited to a small set of history positions.

For a request with $N$ interactions, we sample a set $\mathcal S$ of $\min(\kappa,N)$ positions uniformly without replacement, where $\kappa$ is the per-request re-encoding budget. Sampled positions use the current IME; the others read from the cache:
\begin{equation}
\widetilde Z_t=
\begin{cases}
C_\phi(x_{i_t},e_{i_t}),&t\in\mathcal S,\\
\operatorname{sg}[Z^{\mathrm{cache}}_{i_t}],&t\notin\mathcal S.
\end{cases}
\label{eq:sampled}
\end{equation}
Live memories replace cached blocks at their original positions, preserving history order and length. Retrieval gradients reach the IME through the live blocks. Cached blocks still participate in RE computation and update request-side parameters and the memory input mapping.

Figure~\ref{fig:joint-training} illustrates the gradient paths. The two encoders share a frozen language model backbone with separate LoRA adapters~\cite{lora}. The IME starts from semantic pretraining and updates its summary embeddings and LoRA parameters while fixing the item identity embeddings and projection. The request side jointly learns its LoRA adapter, memory input mapping, REC tokens, and representation adapters.

The joint objective combines final retrieval, segment prediction, and semantic preservation:
\begin{equation}
\begin{aligned}
\mathcal L_{\mathrm{joint}}&=\mathcal L_{\mathrm{ret}}
+\lambda_{\mathrm{seg}}\mathcal L_{\mathrm{seg}}
+\lambda_{\mathrm{sem}}\mathcal L_{\mathrm{sem}},\\
\mathcal L_{\mathrm{ret}}&=\ell(u,y;\mathcal N_{\mathrm{batch}}).
\end{aligned}
\label{eq:stage-b}
\end{equation}
The set $\mathcal N_{\mathrm{batch}}$ contains target items from other requests in the same training group, excluding identities matching the positive. These candidates are encoded live, so the loss optimizes both request and candidate representations. The segment loss is defined in Equation~\eqref{eq:segment}. The semantic term uses the restricted-attention reconstruction cross-entropy from Section~\ref{sec:pretraining}, evaluated only on sampled, re-encoded items. Its weight and $\lambda_{\mathrm{seg}}$ control semantic and sequential supervision.

Encoder updates gradually make cached representations stale. At the end of each training epoch, we re-encode the catalog with the current parameters to refresh both the item memory cache and the trainable candidate cache. Refreshes do not retain a computation graph and use the same item inputs and encoding paths as live computation. Before evaluation, caches are rebuilt from the selected checkpoint.

Dataset-specific training and readout settings are given in Section~\ref{sec:setup}.

\subsection{Computational Savings and Cost Amortization}
\label{sec:cost}
CRIM invests offline computation in memory learning and cache maintenance to reduce repeated online computation. The reference uses the same backbone and encodes full history text per request. Both precompute candidate vectors and retrieve from the same catalog.

\paragraph{Per-request savings.}
For $N$ history items averaging $L$ text tokens each, direct encoding uses $NL+L_0$ positions, where $L_0$ is the query and template length. CRIM uses $K$ vectors per item and $R$ REC tokens per segment, giving
\begin{equation}
\gamma=\frac{NK+JR+L_0}{NL+L_0}.
\label{eq:compression}
\end{equation}
For long histories, $\gamma$ approaches $(K+R/b)/L$. Let $c_{\mathrm{lin}}$ and $c_{\mathrm{attn}}$ denote the reference costs of linear projections/feed-forward computation and attention. Compression scales them approximately by $\gamma$ and $\gamma^2$. With additional memory-input operations costing $c_{\mathrm{add}}$, per-request savings are
\begin{equation}
\Delta c\approx(1-\gamma)c_{\mathrm{lin}}+
(1-\gamma^2)c_{\mathrm{attn}}-c_{\mathrm{add}}.
\label{eq:saving}
\end{equation}
The common candidate-matching cost cancels. Longer item text and fewer memory vectors increase request-encoding savings; total savings also depend on the share of candidate matching and input adaptation.

\paragraph{Cumulative savings.}
For a given serving horizon and refresh schedule, let $\Delta C_{\mathrm{off}}$ denote CRIM's additional offline cost, including pretraining, joint training, initial cache construction, and maintenance. Net savings after $Q$ requests and the break-even volume, for $\Delta C_{\mathrm{off}}>0$ and $\Delta c>0$, are
\begin{align}
S(Q)&=Q\Delta c-\Delta C_{\mathrm{off}},
\label{eq:total-saving}\\
Q^*&=\left\lceil\frac{\Delta C_{\mathrm{off}}}{\Delta c}\right\rceil.
\label{eq:break-even}
\end{align}
For $Q\geq Q^*$, inference savings cover the additional offline investment; each further request saves $\Delta c$. The benefit depends on compression gains and memory reuse within an update cycle. Section~\ref{sec:efficiency} measures these costs on LazSeq.

\begin{table*}[!t]
\centering
\caption{Full-catalog retrieval on LazSeq (histories up to 500 items). Bold/underlined values mark the best/second-best results; dashes denote metrics unavailable from beam-limited outputs.}
\label{tab:lazseq-main}
\setlength{\tabcolsep}{2pt}
\begin{tabular*}{\textwidth}{@{\extracolsep{\fill}}lcrrrrrrrrrr@{}}
\toprule
& & \multicolumn{5}{c}{Recall} & \multicolumn{3}{c}{NDCG} & \multicolumn{2}{c}{MRR} \\
\cmidrule(lr){3-7}\cmidrule(lr){8-10}\cmidrule(lr){11-12}
Method & Year & @10 & @50 & @100 & @1000 & @3000 & @10 & @50 & @100 & @100 & All \\
\midrule
GRU4Rec & 2016 & 0.0300 & 0.0670 & 0.0944 & 0.2598 & 0.3951 & 0.0164 & 0.0246 & 0.0290 & 0.0144 & 0.0151 \\
NARM & 2017 & 0.0223 & 0.0634 & 0.0894 & 0.2544 & 0.3941 & 0.0119 & 0.0207 & 0.0250 & 0.0109 & 0.0116 \\
SASRec & 2018 & 0.0313 & 0.0697 & 0.0937 & 0.2487 & 0.3765 & 0.0170 & 0.0253 & 0.0292 & 0.0147 & 0.0154 \\
STAMP & 2018 & 0.0290 & 0.0620 & 0.0794 & 0.2274 & 0.3588 & 0.0153 & 0.0225 & 0.0253 & 0.0128 & 0.0135 \\
BERT4Rec & 2019 & 0.0250 & 0.0600 & 0.0860 & 0.2661 & 0.4201 & 0.0127 & 0.0204 & 0.0246 & 0.0109 & 0.0117 \\
S3Rec & 2020 & 0.0340 & 0.0794 & 0.1097 & 0.3331 & 0.4965 & 0.0174 & 0.0272 & 0.0321 & 0.0148 & 0.0157 \\
LightSANs & 2021 & 0.0317 & 0.0730 & 0.0960 & 0.2511 & 0.3815 & 0.0163 & 0.0255 & 0.0292 & 0.0139 & 0.0146 \\
SINE & 2021 & 0.0357 & 0.0784 & 0.1030 & 0.2761 & 0.3998 & 0.0199 & 0.0293 & 0.0333 & 0.0174 & 0.0182 \\
FEARec & 2023 & 0.0333 & 0.0677 & 0.0950 & 0.2324 & 0.3738 & 0.0177 & 0.0253 & 0.0297 & 0.0150 & 0.0156 \\
TIGER & 2023 & 0.0317 & 0.0737 & 0.0927 & \textemdash & \textemdash & 0.0159 & 0.0251 & 0.0282 & 0.0133 & \textemdash \\
LlamaRec & 2023 & 0.0290 & 0.0637 & 0.0827 & 0.2381 & 0.3745 & 0.0160 & 0.0233 & 0.0263 & 0.0137 & 0.0144 \\
GenSAR & 2025 & 0.2121 & 0.3768 & 0.4308 & \textemdash & \textemdash & 0.1241 & 0.1608 & 0.1696 & 0.1060 & \textemdash \\
GEMS & 2026 & 0.0100 & 0.0333 & 0.0487 & 0.1627 & 0.3131 & 0.0066 & 0.0116 & 0.0141 & 0.0068 & 0.0073 \\
GTE-1.5B & 2024 & \underline{0.3317} & 0.5783 & \underline{0.6803} & 0.9097 & 0.9630 & \underline{0.1890} & 0.2427 & 0.2593 & \underline{0.1583} & \underline{0.1594} \\
Gemma3-4B-PT & 2025 & 0.3030 & 0.5627 & 0.6657 & 0.9150 & 0.9667 & 0.1744 & 0.2319 & 0.2486 & 0.1489 & 0.1500 \\
Gemma3-4B-IT & 2025 & 0.2897 & 0.5500 & 0.6643 & 0.9140 & 0.9647 & 0.1617 & 0.2191 & 0.2377 & 0.1366 & 0.1377 \\
Llama3.1-8B & 2024 & 0.2983 & 0.5627 & 0.6787 & 0.9227 & 0.9693 & 0.1701 & 0.2285 & 0.2474 & 0.1454 & 0.1466 \\
Ministral-8B-IT & 2024 & 0.3057 & 0.5647 & 0.6757 & 0.9197 & 0.9693 & 0.1721 & 0.2295 & 0.2475 & 0.1452 & 0.1463 \\
Phi-4-14B & 2024 & 0.3153 & 0.5733 & 0.6800 & \underline{0.9253} & \textbf{0.9750} & 0.1790 & 0.2357 & 0.2530 & 0.1510 & 0.1521 \\
Qwen3-Emb.-0.6B & 2025 & 0.2953 & 0.5467 & 0.6447 & 0.8953 & 0.9580 & 0.1745 & 0.2301 & 0.2460 & 0.1508 & 0.1520 \\
Qwen3-Emb.-4B & 2025 & \textbf{0.3390} & \underline{0.5807} & 0.6800 & 0.9177 & 0.9630 & \textbf{0.1923} & \textbf{0.2458} & \textbf{0.2620} & \textbf{0.1604} & \textbf{0.1615} \\
\textbf{CRIM} & Ours & 0.3204 & \textbf{0.5835} & \textbf{0.6969} & \textbf{0.9330} & \underline{0.9737} & 0.1850 & \underline{0.2430} & \underline{0.2614} & 0.1576 & 0.1587 \\
\bottomrule
\end{tabular*}
\end{table*}

\begin{table}[!htbp]
\centering
\caption{Full-catalog retrieval on Qilin (histories up to 100 items). Bold/underlined values mark the best/second-best results; dashes denote metrics unavailable from beam-limited outputs.}
\label{tab:qilin-main}
\setlength{\tabcolsep}{1.2pt}
\resizebox{\columnwidth}{!}{%
\begin{tabular}{@{}lcrrrrrr@{}}
\toprule
& & \multicolumn{2}{c}{Recall} & \multicolumn{2}{c}{NDCG} & \multicolumn{2}{c}{MRR} \\
\cmidrule(lr){3-4}\cmidrule(lr){5-6}\cmidrule(lr){7-8}
Method & Year & @10 & @100 & @10 & @100 & @100 & All \\
\midrule
GRU4Rec & 2016 & 0.0000 & 0.0000 & 0.0000 & 0.0000 & 0.0000 & 0.0001 \\
NARM & 2017 & 0.0000 & 0.0000 & 0.0000 & 0.0000 & 0.0000 & 0.0000 \\
SASRec & 2018 & 0.0000 & 0.0000 & 0.0000 & 0.0000 & 0.0000 & 0.0001 \\
STAMP & 2018 & 0.0034 & 0.0034 & 0.0028 & 0.0028 & 0.0026 & 0.0026 \\
BERT4Rec & 2019 & 0.0000 & 0.0011 & 0.0000 & 0.0002 & 0.0000 & 0.0001 \\
S3Rec & 2020 & 0.0000 & 0.0000 & 0.0000 & 0.0000 & 0.0000 & 0.0000 \\
LightSANs & 2021 & 0.0000 & 0.0011 & 0.0000 & 0.0002 & 0.0000 & 0.0002 \\
SINE & 2021 & 0.0011 & 0.0022 & 0.0005 & 0.0007 & 0.0003 & 0.0004 \\
FEARec & 2023 & 0.0000 & 0.0000 & 0.0000 & 0.0000 & 0.0000 & 0.0001 \\
TIGER & 2023 & 0.0022 & 0.0022 & 0.0022 & 0.0022 & 0.0022 & \textemdash \\
LlamaRec & 2023 & 0.0011 & 0.0034 & 0.0004 & 0.0008 & 0.0002 & 0.0004 \\
GenSAR & 2025 & 0.0101 & 0.0246 & 0.0054 & 0.0084 & 0.0046 & \textemdash \\
GEMS & 2026 & 0.0000 & 0.0000 & 0.0000 & 0.0000 & 0.0000 & 0.0001 \\
GTE-1.5B & 2024 & \textbf{0.5969} & \underline{0.7548} & \underline{0.4404} & \underline{0.4739} & \underline{0.3977} & \underline{0.3981} \\
E5-Mistral-7B & 2023 & 0.5039 & 0.6887 & 0.3542 & 0.3927 & 0.3146 & 0.3151 \\
Gemma3-4B-PT & 2025 & 0.5151 & 0.7021 & 0.3694 & 0.4085 & 0.3319 & 0.3324 \\
Gemma3-4B-IT & 2025 & 0.5017 & 0.6943 & 0.3636 & 0.4042 & 0.3286 & 0.3291 \\
Llama3.1-8B & 2024 & 0.5297 & 0.7044 & 0.3806 & 0.4187 & 0.3419 & 0.3424 \\
Ministral-8B-IT & 2024 & 0.5364 & 0.7055 & 0.3901 & 0.4252 & 0.3506 & 0.3512 \\
Phi-4-14B & 2024 & 0.5409 & 0.7100 & 0.3991 & 0.4355 & 0.3619 & 0.3625 \\
Qwen3-4B & 2025 & 0.5733 & 0.7301 & 0.4052 & 0.4373 & 0.3582 & 0.3587 \\
Qwen3-Emb.-0.6B & 2025 & 0.5465 & 0.7167 & 0.3984 & 0.4339 & 0.3585 & 0.3590 \\
Qwen3-Emb.-4B & 2025 & 0.5801 & \textbf{0.7592} & 0.4126 & 0.4518 & 0.3678 & 0.3682 \\
\textbf{CRIM} & Ours & \underline{0.5845} & 0.7458 & \textbf{0.4434} & \textbf{0.4772} & \textbf{0.4054} & \textbf{0.4059} \\
\bottomrule
\end{tabular}%
}
\end{table}

\begin{table}[!htbp]
\centering
\caption{Full-catalog retrieval on ML1M (histories up to 100 items). Bold/underlined values mark the best/second-best results; dashes denote metrics unavailable from beam-limited outputs.}
\label{tab:ml1m-main}
\setlength{\tabcolsep}{1.2pt}
\resizebox{\columnwidth}{!}{%
\begin{tabular}{@{}lcrrrrrr@{}}
\toprule
& & \multicolumn{2}{c}{Recall} & \multicolumn{2}{c}{NDCG} & \multicolumn{2}{c}{MRR} \\
\cmidrule(lr){3-4}\cmidrule(lr){5-6}\cmidrule(lr){7-8}
Method & Year & @10 & @100 & @10 & @100 & @100 & All \\
\midrule
GRU4Rec & 2016 & 0.0510 & 0.1555 & 0.0269 & 0.0465 & 0.0227 & 0.0244 \\
NARM & 2017 & 0.0510 & 0.1671 & 0.0302 & 0.0523 & 0.0273 & 0.0290 \\
SASRec & 2018 & 0.0534 & 0.1740 & 0.0300 & 0.0531 & 0.0263 & 0.0279 \\
STAMP & 2018 & 0.0534 & 0.1647 & 0.0314 & 0.0531 & 0.0282 & 0.0298 \\
BERT4Rec & 2019 & 0.0487 & 0.2019 & 0.0279 & 0.0565 & 0.0259 & 0.0276 \\
S3Rec & 2020 & 0.0696 & 0.2065 & 0.0367 & 0.0629 & 0.0310 & 0.0330 \\
LightSANs & 2021 & 0.0534 & 0.1601 & 0.0273 & 0.0475 & 0.0221 & 0.0238 \\
SINE & 2021 & 0.0580 & 0.1671 & 0.0346 & 0.0548 & 0.0302 & 0.0318 \\
FEARec & 2023 & 0.0510 & 0.1717 & 0.0268 & 0.0508 & 0.0238 & 0.0254 \\
TIGER & 2023 & 0.0371 & 0.1647 & 0.0215 & 0.0463 & 0.0207 & \textemdash \\
LlamaRec & 2023 & 0.0557 & 0.1856 & 0.0322 & 0.0565 & 0.0287 & 0.0303 \\
GenSAR & 2025 & 0.0673 & 0.2251 & \underline{0.0403} & 0.0689 & \underline{0.0360} & \textemdash \\
GEMS & 2026 & 0.0046 & 0.0650 & 0.0017 & 0.0126 & 0.0023 & 0.0041 \\
GTE-1.5B & 2024 & 0.0580 & \underline{0.2900} & 0.0273 & 0.0698 & 0.0243 & 0.0259 \\
E5-Mistral-7B & 2023 & 0.0487 & 0.2529 & 0.0231 & 0.0619 & 0.0215 & 0.0232 \\
Gemma3-4B-PT & 2025 & 0.0487 & 0.2135 & 0.0223 & 0.0529 & 0.0187 & 0.0204 \\
Gemma3-4B-IT & 2025 & 0.0464 & 0.2088 & 0.0211 & 0.0523 & 0.0181 & 0.0197 \\
Llama3.1-8B & 2024 & 0.0487 & 0.1949 & 0.0272 & 0.0556 & 0.0255 & 0.0272 \\
Ministral-8B-IT & 2024 & \underline{0.0719} & 0.2807 & 0.0315 & 0.0723 & 0.0263 & 0.0282 \\
Phi-4-14B & 2024 & 0.0696 & 0.2854 & 0.0320 & 0.0742 & 0.0279 & 0.0297 \\
Qwen3-4B & 2025 & 0.0557 & 0.2390 & 0.0254 & 0.0608 & 0.0222 & 0.0238 \\
Qwen3-Emb.-0.6B & 2025 & 0.0255 & 0.2251 & 0.0130 & 0.0508 & 0.0151 & 0.0166 \\
Qwen3-Emb.-4B & 2025 & 0.0580 & 0.2831 & 0.0328 & \underline{0.0771} & 0.0327 & \underline{0.0344} \\
\textbf{CRIM} & Ours & \textbf{0.1183} & \textbf{0.3782} & \textbf{0.0629} & \textbf{0.1139} & \textbf{0.0548} & \textbf{0.0564} \\
\bottomrule
\end{tabular}%
}
\end{table}

\section{Experiments}
\label{sec:experiments}

\subsection{Experimental Setup}
\label{sec:setup}

\paragraph{Datasets.}
We evaluate on LazSeq, Qilin, and MovieLens-1M (ML1M). LazSeq and Qilin support query-conditioned personalized retrieval, where a model combines the user history and current query to predict a target item. ML1M supports sequential recommendation without an explicit query. Each request contains one target. History lengths are capped at 500 for LazSeq and 100 for Qilin and ML1M. All ML1M methods use the same request-level data protocol, without expanding history prefixes into additional training examples. Appendix~\ref{app:training} gives dataset statistics, loading details, and checkpoint-selection protocols.

\begin{sloppypar}
\paragraph{Baselines.}
All baseline results are obtained from our own reproductions, and we will release the corresponding code. We compare three groups of methods. The ID-based group contains GRU4Rec~\cite{gru4rec}, NARM~\cite{narm}, STAMP~\cite{stamp}, SASRec~\cite{sasrec}, BERT4Rec~\cite{bert4rec}, S3Rec~\cite{s3rec}, LightSANs~\cite{lightsans}, SINE~\cite{sine}, and FEARec~\cite{fearec}. Generative retrieval and language-model recommendation methods include TIGER~\cite{tiger}, LlamaRec~\cite{llamarec}, GenSAR~\cite{gensar}, and GEMS~\cite{gems}. The dense text encoders include GTE-1.5B~\cite{gteqwen2}, E5-Mistral-7B~\cite{e5mistral}, Gemma3-4B (pretrained and instruction-tuned), Llama3.1-8B, Ministral-8B-Instruct, Phi-4-14B, Qwen3-4B, and Qwen3-Embedding-0.6B and -4B. Qwen3-Embedding-4B shares CRIM's backbone and is the main reference for examining memory compression and computational cost.

ID-based methods use history item IDs. CRIM and query-aware dense encoders receive history item text and the query on LazSeq and Qilin. TIGER, LlamaRec, GenSAR, and GEMS use implementations adapted to the current retrieval tasks. Dense results use the corresponding LoRA-trained checkpoints.

\end{sloppypar}

\paragraph{Evaluation.}
We evaluate retrieval over the complete valid item catalog, retaining items that already occur in the user history. Recall@$K$ and NDCG@$K$ measure target retrieval and ranking quality. Results are averaged across test requests. We additionally report MRR@100, assigning zero to targets ranked below 100, and full-catalog MRR where available. Beam-limited runs without complete ranks do not report full-catalog MRR.

\paragraph{Implementation.}
We train with eight NVIDIA A100-80GB GPUs. CRIM uses Qwen3-Embedding-4B in BF16 with LoRA rank 16 for both encoders. A shared set of $K=4$ summary tokens produces four memory vectors per item. Each history segment contains $b=4$ items followed by $R=4$ REC tokens. The re-encoding budget is $\kappa=8$ on LazSeq and $\kappa=4$ on Qilin and ML1M, and item memories and trainable candidate caches are refreshed after each epoch. Semantic pretraining uses a KL weight of 0.1. Joint training uses a learning rate of $10^{-4}$, temperature 0.05, segment loss weight 1, and semantic reconstruction weight 0.1.

On LazSeq and Qilin, joint training runs for at most three epochs with a global batch size of 64. The retrieval loss is computed within logical groups of at most eight target items. ML1M uses at most ten epochs with batch size eight, a mean readout from the final REC block, fixed pooled candidate encodings, and an additional title-generation loss. Its retrieval negatives also include catalog-sampled items. Table~\ref{tab:trainingconfig} in Appendix~\ref{app:training} gives the full configurations.

\subsection{The LazSeq Dataset}
\label{sec:lazseq}
We construct LazSeq for personalized search with long interaction histories. Each request contains an ordered history, a current query, and a target item; history and target items have shared IDs and titles. Repeated interactions are retained. A shared item catalog supports the study of semantic reuse across requests.

\paragraph{Scale and characteristics.}
LazSeq contains approximately 30,000 requests with up to 500 history interactions each. Its shared catalog comprises 49,993 distinct items from history and target positions. CRIM uses 23,999 training requests, 3,000 validation requests, and 2,999 test requests. Fixed request splits and a title catalog support reproducible comparisons. Figure~\ref{fig:lazseq-construction} in Appendix~\ref{app:lazseq} illustrates query--history filtering and catalog construction; the appendix also details the request splits.

\subsection{Retrieval Results and Comparative Analysis}
\label{sec:results}
Tables~\ref{tab:lazseq-main}--\ref{tab:ml1m-main} compare CRIM with sequential, generative, and dense retrieval methods. We report Recall, NDCG, MRR@100, and full-catalog MRR, with cutoffs shown in the tables. Relative improvements use unrounded results. Appendix~\ref{app:comparisons} compares method groups and additional ranking cutoffs.

\paragraph{Long-history retrieval on LazSeq.}
CRIM achieves the highest Recall@50, Recall@100, and Recall@1000, reaching 0.5835, 0.6969, and 0.9330, respectively. These values improve over the strongest baseline at each cutoff by 0.49\%, 2.43\%, and 0.83\%: Qwen3-Embedding-4B at 50, GTE-1.5B at 100, and Phi-4 at 1000. CRIM ranks second at NDCG@50 and NDCG@100, reaching 0.2430 and 0.2614. At Recall@3000, it ranks second with 0.9737, following Phi-4 at 0.9750. Compared with Qwen3-Embedding-4B, CRIM improves Recall@50, @100, @1000, and @3000 by 0.49\%, 2.49\%, 1.67\%, and 1.11\%, respectively. These results show that four memory vectors per history item preserve the information needed for strong candidate retrieval over long histories.

\paragraph{Query-conditioned retrieval on Qilin.}
CRIM achieves the best NDCG@10, NDCG@100, MRR@100, and full-catalog MRR. Relative to Qwen3-Embedding-4B, NDCG@10 rises from 0.4126 to 0.4434, NDCG@100 from 0.4518 to 0.4772, and full-catalog MRR from 0.3682 to 0.4059, corresponding to gains of 7.48\%, 5.64\%, and 10.23\%. CRIM also exceeds GTE-1.5B, the strongest baseline for these three metrics, by 0.70\%, 0.71\%, and 1.96\%.

\paragraph{Sequential recommendation on ML1M.}
Without an explicit query, CRIM leads all six reported metrics. Its Recall@10 of 0.1183 exceeds the strongest baseline, Ministral-8B-Instruct at 0.0719, by 64.52\%. Recall@100 reaches 0.3782, improving over GTE-1.5B's 0.2900 by 30.40\% and over Qwen3-Embedding-4B by 33.61\%. NDCG@100 reaches 0.1139, a 47.67\% gain over the strongest baseline, Qwen3-Embedding-4B. NDCG@10, MRR@100, and full-catalog MRR are 0.0629, 0.0548, and 0.0564, respectively. These results demonstrate improvements in both candidate retrieval and ranking quality for history-only recommendation under the ML1M configuration.

\subsection{Efficiency and Cost Analysis}
\label{sec:efficiency}
\begin{figure}[!htbp]
  \centering
  \includegraphics[width=\columnwidth]{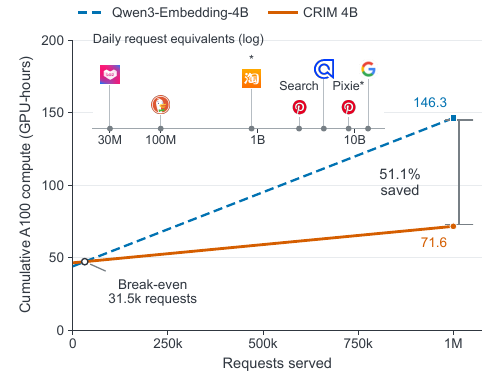}
  \caption{Cumulative A100 compute on LazSeq: optimization, initial cache, and warm-cache inference. Inset: published service volumes~\cite{duckduckgo,rankgr,pinterestsearch,algolia,pixie,googlesearch}; the two P logos denote Pinterest Search and Pixie recommendations. Lazada: 30M searches/day (2020)~\cite{lazadasearch2020}. Stars mark conversions from requests/s to requests/day (Appendix~\ref{app:timing}).}
  \Description{CRIM and Qwen3-Embedding-4B cumulative compute curves cross near 31,500 requests. At one million requests they reach 71.6 and 146.3 GPU-hours, respectively. A separate inset places public services on a logarithmic daily request-equivalent scale. The two Pinterest logos identify Search and Pixie recommendation workloads.}
  \label{fig:cost}
\end{figure}

\paragraph{Online inference.}
Figure~\ref{fig:quality-efficiency} measures end-to-end, batch-one latency on a single A100-SXM4-80GB over 200 shared LazSeq requests. Timing covers input preparation, request encoding, full-catalog retrieval, and returned IDs, with prebuilt caches. Appendix~\ref{app:timing} details warm-ups, repetitions, and cost calibration.

CRIM reduces mean latency to 90.76\,ms from the dense reference's 367.45\,ms, a $4.05\times$ speedup. The 95th-percentile latency falls from 818.03 to 173.39\,ms. Mean serving input length decreases from 5,020.36 to 966.62 positions, including REC tokens, the query, and template tokens, a reduction of 80.75\%. CRIM is also faster than the 0.6B embedding baseline at 117.17\,ms while achieving higher Recall@100. Independent item encoding therefore shifts substantial work out of the online request path.

\paragraph{Cost amortization.}
Figure~\ref{fig:cost} combines estimated A100 optimization and initial cache costs with serial warm-cache inference. The estimated upfront costs are 46.44 GPU-hours for CRIM and 44.02 for Qwen3-Embedding-4B. The online savings recover this difference after approximately 31,500 requests. At one million requests, cumulative compute reaches 71.63 and 146.33 GPU-hours, respectively, yielding a 51.05\% saving. At ten million requests, the saving increases to 72.05\%. This scenario holds the deployed model and catalog fixed. Training-time cache refreshes and later updates add to $\Delta C_{\mathrm{off}}$ (Appendix~\ref{app:timing}).

\subsection{Ablation Studies}
\label{sec:ablation}
\begin{table}[!htbp]
\centering
\caption{LazSeq supervision and pretraining ablations (\%). R, N, and M denote Recall, NDCG, and MRR. Best retrieval metrics are bold.}
\label{tab:objectives}
\small\setlength{\tabcolsep}{4pt}
\begin{tabular*}{\columnwidth}{@{\extracolsep{\fill}}lrrr@{}}
\toprule
Variant & R@100 & N@100 & M@100 \\
\midrule
\textbf{CRIM} & \textbf{69.69} & \textbf{26.14} & \textbf{15.76} \\
w/o $\mathcal L_{\mathrm{seg}}$ & 65.99 & 24.19 & 14.23 \\
w/o $\mathcal L_{\mathrm{sem}}$ & 69.42 & 25.69 & 15.40 \\
Retrieval loss only & 65.66 & 24.15 & 14.23 \\
w/o REC and $\mathcal L_{\mathrm{seg}}$ & 66.29 & 24.39 & 14.39 \\
w/o semantic pretraining & 68.66 & 25.36 & 15.05 \\
\bottomrule
\end{tabular*}
\end{table}

We conduct ablations on held-out LazSeq requests with a 49,993-item catalog. Each run selects its checkpoint by validation Recall@3000. Table~\ref{tab:objectives} reports Recall, NDCG, and MRR at 100.

\paragraph{Sequence supervision.}
CRIM exceeds the variant without segment prediction by 3.70 percentage points in Recall@100, 1.94 points in NDCG@100, and 1.53 points in MRR@100. Retrieval loss alone gives 65.66\% Recall@100; removing REC tokens together with segment supervision gives 66.29\%. Across three seeds, segment supervision improves Recall@100 by 2.53--3.97 percentage points. The variant without semantic reconstruction trails CRIM by 0.27, 0.45, and 0.36 points in these metrics (Appendix~\ref{app:seeds}).

\paragraph{Semantic pretraining.}
CRIM exceeds the variant without semantic pretraining by 1.03 percentage points in Recall@100, 0.77 points in NDCG@100, and 0.71 points in MRR@100. Initializing item memories through text reconstruction therefore benefits subsequent joint training.

\paragraph{Cache-based joint optimization.}
Training-time re-encoding budgets $\kappa=1,4,8$ give Recall@100 between 69.56\% and 69.69\% (Appendix~\ref{app:cache}, Table~\ref{tab:cache}). CRIM exceeds the fixed-cache variant by 3.07 percentage points in Recall@100; refreshing every two epochs gives 68.49\% recall.

\paragraph{Memory capacity.}
We vary $K$ over 1, 4, 8, 16, and 32. Four vectors give the highest Recall@100 and NDCG@100 while using a 2.560\,GB serving cache, compared with 8.703\,GB at $K=16$ and 16.894\,GB at $K=32$ (Appendix~\ref{app:capacity}).

\paragraph{Additional analyses.}
Appendix~\ref{app:live-probe} compares sampled and full-history training costs; Appendix~\ref{app:interventions} reports history and query interventions with paired confidence intervals; Appendix~\ref{app:subgroups} analyzes history-length and item-frequency groups.

\section{Conclusion}
We presented CRIM, a personalized retrieval framework that separates reusable item semantics from request-dependent history modeling. Summary tokens compress items into continuous memories, and REC tokens provide predictive supervision within ordered histories. Cached memories and sampled live re-encoding jointly train the IME and RE with a small differentiable item-encoding budget. Across LazSeq, Qilin, and ML1M, CRIM achieves strong candidate recall and ranking quality. On LazSeq, it reduces serving input length by 80.75\% and delivers a $4.05\times$ latency speedup over the same-scale dense reference.

\label{end:maintext}
\bibliographystyle{ACM-Reference-Format}
\bibliography{sample-base}

@article{movielens,
  author = {Harper, F. Maxwell and Konstan, Joseph A.},
  title = {{The MovieLens Datasets: History and Context}},
  journal = {ACM Transactions on Interactive Intelligent Systems},
  volume = {5},
  number = {4},
  articleno = {19},
  numpages = {19},
  year = {2015},
  doi = {10.1145/2827872},
  url = {https://doi.org/10.1145/2827872}
}

@inproceedings{gru4rec,
  author = {Hidasi, Bal{\'a}zs and Karatzoglou, Alexandros and Baltrunas, Linas and Tikk, Domonkos},
  title = {{Session-based Recommendations with Recurrent Neural Networks}},
  year = {2016},
  booktitle = {International Conference on Learning Representations},
  url = {https://arxiv.org/abs/1511.06939}
}

@inproceedings{narm,
  author = {Li, Jing and Ren, Pengjie and Chen, Zhumin and Ren, Zhaochun and Lian, Tao and Ma, Jun},
  title = {{Neural Attentive Session-based Recommendation}},
  booktitle = {Proceedings of the 2017 ACM on Conference on Information and Knowledge Management},
  year = {2017},
  pages = {1419--1428},
  doi = {10.1145/3132847.3132926},
  url = {https://doi.org/10.1145/3132847.3132926}
}

@inproceedings{sasrec,
  author = {Kang, Wang-Cheng and McAuley, Julian},
  title = {{Self-Attentive Sequential Recommendation}},
  year = {2018},
  booktitle = {2018 IEEE International Conference on Data Mining},
  url = {https://arxiv.org/abs/1808.09781}
}

@inproceedings{stamp,
 author = {Liu, Qiao and Zeng, Yifu and Mokhosi, Refuoe and Zhang, Haibin},
 title = {STAMP: Short-Term Attention/Memory Priority Model for Session-based Recommendation},
 booktitle = {Proceedings of the 24th ACM SIGKDD International Conference on Knowledge Discovery \& Data Mining},
 year = {2018},
 location = {London, United Kingdom},
 pages = {1831--1839},
}

@inproceedings{bert4rec,
  author = {Sun, Fei and Liu, Jun and Wu, Jian and Pei, Changhua and Lin, Xiao and Ou, Wenwu and Jiang, Peng},
  title = {{BERT4Rec: Sequential Recommendation with Bidirectional Encoder Representations from Transformer}},
  year = {2019},
  booktitle = {Proceedings of the 28th ACM International Conference on Information and Knowledge Management},
  url = {https://arxiv.org/abs/1904.06690}
}

@inproceedings{s3rec,
  author    = {Kun Zhou and
               Hui Wang and
               Wayne Xin Zhao and
               Yutao Zhu and
               Sirui Wang and
               Fuzheng Zhang and
               Zhongyuan Wang and
               Ji{-}Rong Wen},
  title     = {S3-Rec: Self-Supervised Learning for Sequential Recommendation with
               Mutual Information Maximization},
  booktitle = {{CIKM} '20: The 29th {ACM} International Conference on Information
               and Knowledge Management, Virtual Event, Ireland, October 19-23, 2020},
  pages     = {1893--1902},
  publisher = {{ACM}},
  year      = {2020}
}

@inproceedings{lightsans,
    title  = {Lighter and Better: Low-Rank Decomposed Self-Attention Networks for Next-Item Recommendation},
    author = {Xinyan Fan and
              Zheng Liu and
              Jianxun Lian and
              Wayne Xin Zhao and
              Xing Xie and 
              Ji{-}Rong Wen},
    booktitle = {{SIGIR} '21: The 44th International {ACM} {SIGIR} Conference on Research
               and Development in Information Retrieval, Virtual Event, Canada, July
               11-15, 2021},
    year = {2021},
    pages     = {1733--1737},
    publisher = {{ACM}},
    doi       = {10.1145/3404835.3462978}
}

@inproceedings{sine,
  title={Sparse-interest network for sequential recommendation},
  author={Tan, Qiaoyu and Zhang, Jianwei and Yao, Jiangchao and Liu, Ninghao and Zhou, Jingren and Yang, Hongxia and Hu, Xia},
  booktitle={Proceedings of the 14th ACM International Conference on Web Search and Data Mining},
  pages={598--606},
  year={2021}
}

@inproceedings{fearec,
  author = {Du, Xinyu and Yuan, Huanhuan and Zhao, Pengpeng and Qu, Jianfeng and Zhuang, Fuzhen and Liu, Guanfeng and Liu, Yanchi and Sheng, Victor S.},
  title = {{Frequency Enhanced Hybrid Attention Network for Sequential Recommendation}},
  booktitle = {Proceedings of the 46th International ACM SIGIR Conference on Research and Development in Information Retrieval},
  year = {2023},
  pages = {78--88},
  doi = {10.1145/3539618.3591689},
  url = {https://doi.org/10.1145/3539618.3591689}
}

@inproceedings{tiger,
 author = {Rajput, Shashank and Mehta, Nikhil and Singh, Anima and Hulikal Keshavan, Raghunandan and Vu, Trung and Heldt, Lukasz and Hong, Lichan and Tay, Yi and Tran, Vinh and Samost, Jonah and Kula, Maciej and Chi, Ed and Sathiamoorthy, Maheswaran},
 booktitle = {Advances in Neural Information Processing Systems},
 doi = {10.52202/075280-0452},
 editor = {A. Oh and T. Naumann and A. Globerson and K. Saenko and M. Hardt and S. Levine},
 pages = {10299--10315},
 publisher = {Curran Associates, Inc.},
 title = {Recommender Systems with Generative Retrieval},
 url = {https://proceedings.neurips.cc/paper_files/paper/2023/file/20dcab0f14046a5c6b02b61da9f13229-Paper-Conference.pdf},
 volume = {36},
 year = {2023}
}

@article{llamarec,
  title={LlamaRec: Two-Stage Recommendation using Large Language Models for Ranking},
  author={Yue, Zhenrui and Rabhi, Sara and Moreira, Gabriel de Souza Pereira and Wang, Dong and Oldridge, Even},
  journal={arXiv preprint arXiv:2311.02089},
  year={2023}
}

@inproceedings{gensar,
  author = {Shi, Teng and Xu, Jun and Zhang, Xiao and Zang, Xiaoxue and Zheng, Kai and Song, Yang and Yu, Enyun},
  title = {{GenSAR: Unifying Balanced Search and Recommendation with Generative Retrieval}},
  booktitle = {Proceedings of the Nineteenth ACM Conference on Recommender Systems},
  year = {2025},
  pages = {124--134},
  doi = {10.1145/3705328.3748071},
  url = {https://doi.org/10.1145/3705328.3748071}
}

@inproceedings{gems,
  author = {Zhao, Jujia and Wang, Zihan and Pan, Shuaiqun and Verberne, Suzan and Ren, Zhaochun},
  title = {{Unifying Search and Recommendation in LLMs via Gradient Multi-Subspace Tuning}},
  year = {2026},
  booktitle = {Proceedings of the 49th International ACM SIGIR Conference on Research and Development in Information Retrieval},
  url = {https://arxiv.org/abs/2601.09496}
}

@article{qwen3embedding,
  author = {Zhang, Yanzhao and Li, Mingxin and Long, Dingkun and Zhang, Xin and Lin, Huan and Yang, Baosong and Xie, Pengjun and Yang, An and Liu, Dayiheng and Lin, Junyang and Huang, Fei and Zhou, Jingren},
  title = {{Qwen3 Embedding: Advancing Text Embedding and Reranking Through Foundation Models}},
  year = {2025},
  journal = {arXiv preprint arXiv:2506.05176},
  url = {https://arxiv.org/abs/2506.05176}
}

@misc{gteqwen2,
  author = {{Alibaba-NLP}},
  title = {{gte-Qwen2-1.5B-instruct: Model Card}},
  year = {2024},
  url = {https://huggingface.co/Alibaba-NLP/gte-Qwen2-1.5B-instruct},
  note = {Accessed 22 October 2026}
}

@inproceedings{icae,
  author = {Ge, Tao and Hu, Jing and Wang, Lei and Wang, Xun and Chen, Si-Qing and Wei, Furu},
  title = {{In-context Autoencoder for Context Compression in a Large Language Model}},
  year = {2024},
  booktitle = {The Twelfth International Conference on Learning Representations},
  url = {https://arxiv.org/abs/2307.06945}
}

@inproceedings{recformer,
  author = {Li, Jiacheng and Wang, Ming and Li, Jin and Fu, Jinmiao and Shen, Xin and Shang, Jingbo and McAuley, Julian},
  title = {{Text Is All You Need: Learning Language Representations for Sequential Recommendation}},
  booktitle = {Proceedings of the 29th ACM SIGKDD Conference on Knowledge Discovery and Data Mining},
  year = {2023},
  pages = {1258--1267},
  doi = {10.1145/3580305.3599519},
  url = {https://doi.org/10.1145/3580305.3599519}
}

@inproceedings{ebmc,
    title = "Pretraining Context Compressor for Large Language Models with Embedding-Based Memory",
    author = "Dai, Yuhong  and
      Lian, Jianxun  and
      Huang, Yitian  and
      Zhang, Wei  and
      Zhou, Mingyang  and
      Wu, Mingqi  and
      Xie, Xing  and
      Liao, Hao",
    editor = "Che, Wanxiang  and
      Nabende, Joyce  and
      Shutova, Ekaterina  and
      Pilehvar, Mohammad Taher",
    booktitle = "Proceedings of the 63rd Annual Meeting of the {A}ssociation for {C}omputational {L}inguistics (Volume 1: Long Papers)",
    month = jul,
    year = "2025",
    address = "Vienna, Austria",
    publisher = "Association for Computational Linguistics",
    url = "https://aclanthology.org/2025.acl-long.1394/",
    doi = "10.18653/v1/2025.acl-long.1394",
    pages = "28715--28732",
    ISBN = "979-8-89176-251-0"
}

@article{tokenfactory,
  author = {Chen, Xilun and Wang, Shao-Chuan and Cakici, Baykal and Heldt, Lukasz and Hong, Lichan and Keshavan, Raghu and Nath, Aniruddh and Wei, Li and Yi, Xinyang},
  title = {{Token Factory: Efficiently Integrating Diverse Signals into Large Recommendation Models}},
  year = {2026},
  journal = {arXiv preprint arXiv:2606.19635},
  url = {https://arxiv.org/abs/2606.19635}
}

@article{dualpurpose,
  author = {Li, Baolei and Yuan, Yiping and Zheng, Yilin and Yin, Likang and Liu, Ling and Soldo, Fabio and Rosales, Romer and Yi, Xinyang and Hong, Lichan},
  title = {{Tokens are All You Need: Dual-purpose Semantic IDs for Achieving LLM-Level I/O Efficiency in recommendation systems}},
  year = {2026},
  journal = {arXiv preprint arXiv:2607.24865},
  url = {https://arxiv.org/abs/2607.24865}
}

@inproceedings{gnnautoscale,
  author = {Fey, Matthias and Lenssen, Jan E. and Weichert, Frank and Leskovec, Jure},
  title = {{GNNAutoScale: Scalable and Expressive Graph Neural Networks via Historical Embeddings}},
  booktitle = {Proceedings of the 38th International Conference on Machine Learning},
  pages = {3294--3304},
  year = {2021},
  volume = {139},
  series = {Proceedings of Machine Learning Research},
  publisher = {PMLR},
  url = {https://proceedings.mlr.press/v139/fey21a.html}
}

@inproceedings{lora,
title={Lo{RA}: Low-Rank Adaptation of Large Language Models},
author={Edward J Hu and Yelong Shen and Phillip Wallis and Zeyuan Allen-Zhu and Yuanzhi Li and Shean Wang and Lu Wang and Weizhu Chen},
booktitle={International Conference on Learning Representations},
year={2022},
url={https://openreview.net/forum?id=nZeVKeeFYf9}
}

@misc{duckduckgo,
  author = {{DuckDuckGo}},
  title = {{DuckDuckGo Press Page}},
  year = {n.d.},
  url = {https://duckduckgo.com/press},
  note = {Accessed 22 October 2026}
}

@article{rankgr,
  author = {Fu, Kairui and Wu, Changfa and Yuan, Kun and Cao, Binbin and Huang, Dunxian and Yan, Yuliang and Zheng, Junjun and Zhang, Jianning and Zhou, Silu and Wu, Jian and Kuang, Kun},
  title = {{RankGR: Rank-Enhanced Generative Retrieval with Listwise Direct Preference Optimization in Recommendation}},
  year = {2026},
  journal = {arXiv preprint arXiv:2602.08575},
  url = {https://arxiv.org/abs/2602.08575}
}

@misc{pinterestsearch,
  author = {{Pinterest}},
  title = {{Pinterest Unveils Visual Search Ads and New Performance Tools}},
  year = {2026},
  month = {September},
  url = {https://newsroom.pinterest.com/en-gb/news/new-pinterest-visual-search-performance-ads/},
  note = {17 September 2026; accessed 22 October 2026}
}

@misc{algolia,
  author = {{Algolia}},
  title = {{Algolia Launches Breakthrough Trifecta of AI-Powered Ranking Capabilities}},
  year = {2025},
  month = {March},
  url = {https://www.algolia.com/about/news/algolia-launches-breakthrough-trifecta-of-ai-powered-ranking-capabilities},
  note = {4 March 2025; accessed 22 October 2026}
}

@inproceedings{pixie,
  author = {Eksombatchai, Chantat and Jindal, Pranav and Liu, Jerry Zitao and Liu, Yuchen and Sharma, Rahul and Sugnet, Charles and Ulrich, Mark and Leskovec, Jure},
  title = {{Pixie: A System for Recommending 3+ Billion Items to 200+ Million Users in Real-Time}},
  booktitle = {Proceedings of the 2018 World Wide Web Conference on World Wide Web - WWW '18},
  year = {2018},
  pages = {1775--1784},
  doi = {10.1145/3178876.3186183},
  url = {https://doi.org/10.1145/3178876.3186183}
}

@misc{googlesearch,
  author = {Srinivasan, Vidhya},
  title = {{AI, Personalization and the Future of Shopping}},
  year = {2025},
  month = {March},
  url = {https://blog.google/products/ads-commerce/ai-personalization-and-the-future-of-shopping/},
  note = {3 March 2025; accessed 22 October 2026}
}

@article{e5mistral,
  author = {Wang, Liang and Yang, Nan and Huang, Xiaolong and Yang, Linjun and Majumder, Rangan and Wei, Furu},
  title = {{Improving Text Embeddings with Large Language Models}},
  journal = {arXiv preprint arXiv:2401.00368},
  year = {2024},
  url = {https://arxiv.org/abs/2401.00368}
}

@inproceedings{gst,
  author = {Cao, Kaidi and Phothilimthana, Mangpo and Abu-El-Haija, Sami and Zelle, Dustin and Zhou, Yanqi and Mendis, Charith and Leskovec, Jure and Perozzi, Bryan},
  title = {Learning Large Graph Property Prediction via Graph Segment Training},
  booktitle = {Advances in Neural Information Processing Systems},
  volume = {36},
  year = {2023},
  url = {https://papers.neurips.cc/paper_files/paper/2023/hash/48f8143cebe113f4596e1781771578cd-Abstract-Conference.html}
}

@inproceedings{rest,
  author = {Xue, Rui and Zhao, Tong and Shah, Neil and Liu, Xiaorui},
  title = {Haste Makes Waste: A Simple Approach for Scaling Graph Neural Networks},
  booktitle = {Proceedings of the 42nd International Conference on Machine Learning},
  volume = {267},
  series = {Proceedings of Machine Learning Research},
  pages = {70160--70182},
  publisher = {PMLR},
  year = {2025},
  url = {https://proceedings.mlr.press/v267/xue25e.html}
}

@inproceedings{qilin, series={SIGIR '25}, title={Qilin: A Multimodal Information Retrieval Dataset with APP-level User Sessions}, url={http://dx.doi.org/10.1145/3726302.3730279}, DOI={10.1145/3726302.3730279}, booktitle={Proceedings of the 48th International ACM SIGIR Conference on Research and Development in Information Retrieval}, publisher={ACM}, author={Chen, Jia and Dong, Qian and Li, Haitao and He, Xiaohui and Gao, Yan and Cao, Shaosheng and Wu, Yi and Yang, Ping and Xu, Chen and Hu, Yao and Ai, Qingyao and Liu, Yiqun}, year={2025}, month=jul, pages={3670--3680}, collection={SIGIR '25} }

@misc{lazadasearch2020,
 author = {{MARKETECH APAC}},
 title = {{Lazada PH introduces new solutions to help brands and online sellers}},
 year = {2020},
 month = jul,
 howpublished = {MARKETECH APAC},
 url = {https://marketech-apac.com/lazada-ph-introduces-new-solutions-to-help-brands-and-online-sellers/},
 note = {Published July 1, 2020. Accessed October 26, 2026}
}

@inproceedings{agentdet2026,
  author = {Li, Haolin and Wang, Yaohua and Yan, Ze and Wen, Lijie and Huang, Biqing},
  title = {{AgentDet}: A Shared-Blackboard Multi-Agent Framework for Zero-/Few-Shot Object Detection},
  booktitle = {Proceedings of the IEEE/CVF Conference on Computer Vision and Pattern Recognition (CVPR)},
  month = jun,
  year = {2026},
  pages = {41626--41635}
}

@inproceedings{lira2026,
  title = {Toward Robust Multilingual Adaptation of {LLMs} for Low-Resource Languages},
  author = {Li, Haolin and Zhang, Haipeng and Li, Mang and Wang, Yaohua and Wen, Lijie and Yu, Zhang and Huang, Biqing},
  booktitle = {Proceedings of the 43rd International Conference on Machine Learning},
  pages = {72401--72427},
  year = {2026},
  volume = {306},
  series = {Proceedings of Machine Learning Research},
  publisher = {PMLR},
  url = {https://proceedings.mlr.press/v306/li26jh.html}
}

@inproceedings{prefixtuning,
  title = {Prefix-Tuning: Optimizing Continuous Prompts for Generation},
  author = {Li, Xiang Lisa and Liang, Percy},
  booktitle = {Proceedings of the 59th Annual Meeting of the Association for Computational Linguistics and the 11th International Joint Conference on Natural Language Processing (Volume 1: Long Papers)},
  year = {2021},
  publisher = {Association for Computational Linguistics},
  doi = {10.18653/v1/2021.acl-long.353},
  pages = {4582--4597}
}

@inproceedings{prompttuning,
  title = {The Power of Scale for Parameter-Efficient Prompt Tuning},
  author = {Lester, Brian and Al-Rfou, Rami and Constant, Noah},
  booktitle = {Proceedings of the 2021 Conference on Empirical Methods in Natural Language Processing},
  year = {2021},
  publisher = {Association for Computational Linguistics},
  doi = {10.18653/v1/2021.emnlp-main.243},
  pages = {3045--3059}
}

@inproceedings{llmlingua,
  title = {{LLMLingua}: Compressing Prompts for Accelerated Inference of Large Language Models},
  author = {Jiang, Huiqiang and Wu, Qianhui and Lin, Chin-Yew and Yang, Yuqing and Qiu, Lili},
  booktitle = {Proceedings of the 2023 Conference on Empirical Methods in Natural Language Processing},
  year = {2023},
  publisher = {Association for Computational Linguistics},
  doi = {10.18653/v1/2023.emnlp-main.825},
  pages = {13358--13376}
}
\appendix
\begin{figure*}[!t]
  \centering
  \includegraphics[width=\textwidth]{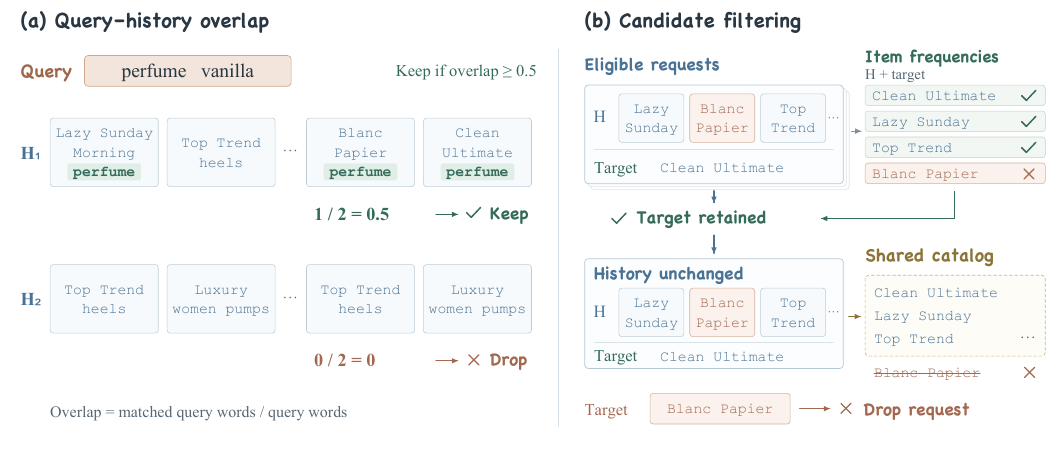}
  \caption{LazSeq construction. (a) Requests are filtered by query--history lexical overlap. (b) Item frequencies in history and target positions determine the retained candidate set; requests are kept when their target is retained.}
  \Description{An example compares histories that meet or fail query overlap filtering. A second panel shows item frequency aggregation, target-based request filtering, and construction of a shared item catalog.}
  \label{fig:lazseq-construction}
\end{figure*}

\section{Implementation and Evaluation Details}
\label{app:protocol}

\subsection{LazSeq Construction}
\label{app:lazseq}
Figure~\ref{fig:lazseq-construction} summarizes the filtering and catalog construction steps.
\paragraph{Request filtering.}
We retain requests with at least 100 original interactions and a lexical query--history overlap of at least 0.5, keeping the most recent 500 history items. To compute overlap, the query is lowercased and split on whitespace, discarding tokens of length one. The score is the fraction of retained query tokens appearing in the concatenated recent history titles. Requests may target either an item already present in the history or a previously unseen item.

\paragraph{Catalog construction and splitting.}
We aggregate eligible requests and count item occurrences in target and retained history positions. These frequencies determine the retained item set; requests are kept if their target belongs to this set. We then fill training, validation, and test splits in processed stream order at an approximate $8:1:1$ ratio. Item frequencies are computed before splitting. The final shared catalog is the union of retained item IDs appearing in the three output splits. This request-level protocol provides a common candidate space across splits.

\subsection{Training and Evaluation Protocols}
\label{app:training}
\begin{table*}[!t]
\centering
\caption{CRIM training configurations for the main results.}
\label{tab:trainingconfig}
\footnotesize\setlength{\tabcolsep}{6pt}
\begin{tabular}{llll}
\toprule
Setting & LazSeq & Qilin & ML1M \\
\midrule
Backbone & \multicolumn{3}{c}{Qwen3-Embedding-4B; bf16; LoRA rank 16} \\
Slots / live history positions & $K=4$ / at most 8 & $K=4$ / at most 4 & $K=4$ / at most 4 \\
Segment length / cache refresh & \multicolumn{3}{c}{4 items / every epoch} \\
Joint training learning rate & \multicolumn{3}{c}{$10^{-4}$; weight decay 0; gradient clip 1} \\
Joint training maximum epochs / batch & 3 / 64 & 3 / 64 & 10 / 8 \\
Selected epoch & 3 & 2 & 7 \\
Validation metric & R@3000 & R@3000 & R@10 \\
Request readout & Suffix EOS & Suffix EOS & Final REC-block mean \\
Candidate tower & Trainable, last token & Trainable, last token & Frozen, pooled store \\
Retrieval candidates during training & 8 targets per logical group & 8 targets per logical group & Positive + cached negatives \\
Segment negatives & \multicolumn{3}{c}{Up to 1,024 catalog negatives} \\
Loss weights (ret., seg., sem., gen.) & $(1,1,0.1,0)$ & $(1,1,0.1,0)$ & $(1,1,0.1,1)$ \\
Query / position & Yes / after history & Yes / after history & No \\
\bottomrule
\end{tabular}
\end{table*}

\paragraph{Model configuration.}
Table~\ref{tab:trainingconfig} gives the training settings. Figures~\ref{fig:slot-learning} and~\ref{fig:joint-training} detail semantic pretraining and joint training. The backbone has hidden dimension 2,560. LazSeq semantic pretraining runs for five epochs with 32 items per GPU, summary learning rate $3\times10^{-3}$, identity learning rate $5\times10^{-4}$, cosine decay, and 5\% warm-up. A 128-dimensional identity embedding projects to two prefix tokens; item titles are capped at 96 tokens. Each segment uses up to 1,024 catalog negatives. Joint training uses weight decay zero and gradient clipping at one. The LazSeq ablations use eight GPUs with the same global batch and logical contrastive groups. The cache-refresh variants refresh candidate vectors every epoch.

LazSeq and Qilin use the EOS readout and joint objective above. The queryless ML1M configuration uses the mean hidden state of the final REC block for the request representation and fixed pooled candidate encodings. It additionally uses a target-title generation loss and catalog-sampled retrieval negatives. Section~\ref{sec:setup} summarizes these settings.

\paragraph{Data and checkpoint protocols.}
Qilin has 7,144/893/893 training/validation/test requests and 159,477 catalog items; ML1M has 3,437/429/431 requests and 3,566 items. Each dataset uses a shared title catalog. LazSeq counts are given in Section~\ref{sec:lazseq}. LazSeq is split at the request level, and a user can occur in multiple splits. Catalog membership is determined before the request split, while model training uses training requests. Filtering the retained catalog may shorten a history below the original 100-interaction eligibility threshold.

The selected CRIM epochs are 3, 2, and 7 for LazSeq, Qilin, and ML1M. Checkpoint selection uses validation Recall@3000, Recall@3000, and Recall@10, respectively. LazSeq ablations select validation Recall@3000, then evaluate the selected checkpoint once on the test set. Full MRR is unavailable for the beam-limited TIGER and GenSAR runs. Method years in the tables indicate publication or specific model release.

\subsection{A100 Measurements and Cost Calibration}
\label{app:timing}
Figure~\ref{fig:quality-efficiency} measures five warm-ups followed by two passes over 200 fixed requests at batch size one. The cache is built before timing. Serving length counts actual RE input positions, including memory slots, REC tokens, query, and special tokens. The request path includes input preparation, encoding, cosine retrieval over 49,993 candidates, and returned IDs. Backbone operations use BF16; the stored serving memories and candidate vectors use FP32. The serving cache occupies 2.560\,GB for CRIM, comprising 2.048\,GB of item memories and 0.512\,GB of candidate vectors. Dense reference candidates occupy 0.512\,GB.

Figure~\ref{fig:cost} uses A100 measurements: estimated optimizer costs are 45.66 and 43.96 GPU-hours for CRIM and the dense reference, and initial cache construction takes 2,809.16 and 206.64 seconds. The online costs are 90.66 and 368.32\,ms per request. These estimates cover optimizer computation, initial cache construction, and inference with a fixed model and catalog. For a cache-refresh or model-update schedule, the corresponding measured maintenance cost enters $\Delta C_{\mathrm{off}}$ in Equation~\eqref{eq:total-saving}. Each additional GPU-hour of relative offline cost requires approximately 12,965 further requests to amortize at these request costs.

For the service-volume inset in Figure~\ref{fig:cost}, monthly and annual totals are divided by 30 and 365, respectively. Starred per-second rates are multiplied by $24\times60\times60=86{,}400$, the number of seconds in a day. They represent daily equivalents at the reported rates. Lazada's approximately 30 million searches per day is a published 2020 figure~\cite{lazadasearch2020}.

\subsection{Additional Comparisons}
\label{app:comparisons}
The method groups differ substantially. Among ID-based models, S3Rec has the highest Recall@100 at 0.1097. The adapted GenSAR implementation reaches 0.4308, Qwen3-Embedding-0.6B reaches 0.6447, and CRIM reaches 0.6969. Methods using item text generally perform better in this query-conditioned setting.

On LazSeq, CRIM ranks second at NDCG@50 and NDCG@100 with 0.2430 and 0.2614, compared with 0.2458 and 0.2620 for Qwen3-Embedding-4B. The dense reference leads Recall@10 at 0.3390 versus 0.3204 for CRIM. CRIM combines higher candidate recall and strong ranking quality with compact history inputs.

On Qilin, Recall@10 improves over Qwen3-Embedding-4B from 0.5801 to 0.5845, while GTE-1.5B achieves 0.5969. CRIM's Recall@100 is 0.7458, compared with 0.7592 for Qwen3-Embedding-4B. Its main benefit on Qilin is therefore better target ranking, complementing the broader-cutoff recall gains observed on LazSeq.

\section{Additional Ablations and Diagnostics}
\label{app:diagnostics}

\subsection{Consistency across Random Seeds}
\label{app:seeds}
\begin{table}[!htbp]
\centering
\caption{Objective ablations across seeds 42, 43, and 44. Values are mean $\pm$ sample standard deviation, in percentages. Best means are bold.}
\label{tab:seeds}
\small\setlength{\tabcolsep}{2pt}
\begin{tabular*}{\columnwidth}{@{\extracolsep{\fill}}lrrr@{}}
\toprule
Variant & R@100 & N@100 & M@100 \\
\midrule
CRIM & $\mathbf{69.68}\pm1.40$ & $\mathbf{26.02}\pm0.20$ & $\mathbf{15.66}\pm0.33$ \\
w/o segment loss & $66.29\pm0.73$ & $24.10\pm0.16$ & $14.10\pm0.12$ \\
w/o semantic loss & $69.12\pm1.11$ & $25.61\pm0.32$ & $15.40\pm0.13$ \\
Retrieval loss only & $66.43\pm0.69$ & $24.12\pm0.03$ & $14.07\pm0.15$ \\
\bottomrule
\end{tabular*}
\end{table}

Table~\ref{tab:seeds} summarizes seeds 42, 43, and 44. Segment supervision improves all three metrics in every seed. Mean Recall@100 is 69.68\% for CRIM, 66.29\% without segment supervision, and 66.43\% with retrieval loss alone. Omitting semantic reconstruction lowers mean Recall@100 by 0.56 and NDCG@100 by 0.41 percentage points, and mean MRR@100 by 0.26 points. The main retrieval benefit of the auxiliary objectives comes from segment supervision.

\subsection{Live Re-encoding and Cache Refresh}
\label{app:cache}
\begin{table}[!htbp]
\centering
\caption{Training-time history re-encoding and item-memory cache refresh on LazSeq. Metrics are percentages; evaluation uses rebuilt caches. Best retrieval metrics are bold.}
\label{tab:cache}
\small\setlength{\tabcolsep}{4pt}
\begin{tabular*}{\columnwidth}{@{\extracolsep{\fill}}lrrr@{}}
\toprule
Variant & R@100 & N@100 & M@100 \\
\midrule
\textbf{CRIM} ($\kappa=8$, epoch refresh) & \textbf{69.69} & \textbf{26.14} & \textbf{15.76} \\
Re-encoding $\kappa=1$ & 69.56 & 25.99 & 15.60 \\
Re-encoding $\kappa=4$ & 69.66 & 25.89 & 15.45 \\
Refresh every two epochs & 68.49 & 25.49 & 15.21 \\
No item-memory cache refresh & 66.62 & 24.70 & 14.70 \\
\bottomrule
\end{tabular*}
\end{table}

Table~\ref{tab:cache} varies training-time re-encoding and item-memory refresh. Recall@100 is 69.56\%, 69.66\%, and 69.69\% for $\kappa=1,4,8$, respectively. CRIM leads all three metrics, reaching 26.14\% NDCG@100 and 15.76\% MRR@100. It exceeds the fixed-cache variant by 3.07 points in Recall@100. Candidate caches refresh each epoch; evaluation uses rebuilt caches.

\subsection{Memory Capacity and Storage}
\label{app:capacity}
\begin{table}[!htbp]
\centering
\caption{Memory capacity on LazSeq. R and N use cutoff 100; M is full-catalog MRR. Metrics are percentages. Cache includes FP32 item memories and candidate vectors. Best retrieval metrics are bold.}
\label{tab:capacity}
\small\setlength{\tabcolsep}{3pt}
\begin{tabular*}{\columnwidth}{@{\extracolsep{\fill}}lrrrr@{}}
\toprule
Variant & R@100 & N@100 & M (all) & Cache (GB) \\
\midrule
$K=1$ & 69.02 & 25.38 & 15.09 & 1.024 \\
CRIM ($K=4$) & \textbf{69.69} & \textbf{26.14} & \textbf{15.87} & 2.560 \\
$K=8$ & 68.22 & 25.20 & 15.07 & 4.607 \\
$K=16$ & 69.22 & 25.75 & 15.49 & 8.703 \\
$K=32$ & 68.20 & 25.31 & 15.13 & 16.894 \\
\bottomrule
\end{tabular*}
\end{table}

Table~\ref{tab:capacity} compares $K=1,4,8,16,32$. CRIM ($K=4$) leads all three metrics. Increasing $K$ to 16 gives 69.22\% Recall@100, while $K=32$ gives 68.20\%. Increasing $K$ beyond four does not improve retrieval in these experiments. For 200 sampled requests, mean RE lengths for $K=1,4,8,16$ are 459.93, 1,126.02, 2,014.14, and 3,790.38 positions.

With hidden dimension $d=2{,}560$ and catalog size $|\mathcal I|=49{,}993$, FP32 serving storage is $4|\mathcal I|d(K+1)$ bytes: $K$ memory vectors and one candidate vector per item. Thus, increasing $K$ from 4 to 16 multiplies storage by 3.4 and mean input length by 3.37 without improving quality. At $K=32$, storage reaches 16.894\,GB. These results support using four summary tokens per item.

\subsection{Live Item Encoding}
\label{app:live-probe}
In a controlled probe on two A100 80GB GPUs, the global batch is 64 and contrastive groups contain eight target items. After one warm-up, three optimizer steps with $\kappa=4$ average 68.45 seconds and allocate at most 23.35\,GB on either GPU. The all-history setting uses up to 500 live positions and runs out of memory during warm-up. The comparison holds the batch, RE history, and loss configuration fixed.

\subsection{History and Query Interventions}
\label{app:interventions}
\begin{table}[!htbp]
\centering
\caption{Inference-time input interventions on LazSeq. Values are percentages. Best values within each group are bold.}
\label{tab:diagnostics}
\small\setlength{\tabcolsep}{4pt}
\begin{tabular*}{\columnwidth}{@{\extracolsep{\fill}}lrrr@{}}
\toprule
Variant & R@100 & N@100 & M@100 \\
\midrule
\multicolumn{4}{l}{History content} \\
Original history & \textbf{69.52} & \textbf{25.93} & \textbf{15.53} \\
Cross-user history swap & 66.12 & 23.67 & 13.62 \\
No history & 65.82 & 23.69 & 13.68 \\
Mean item memories & 64.39 & 23.23 & 13.45 \\
\multicolumn{4}{l}{History order, length, and query} \\
Original history & \textbf{69.69} & \textbf{25.97} & \textbf{15.55} \\
Permutation, seed 42 & 69.56 & 25.62 & 15.18 \\
Permutation, seed 43 & 69.49 & 25.50 & 15.04 \\
Permutation, seed 44 & 69.02 & 25.45 & 15.07 \\
Reverse order & 69.46 & 25.62 & 15.24 \\
Most recent 50 & 68.82 & 25.56 & 15.19 \\
Most recent 100 & 69.19 & 25.76 & 15.37 \\
Most recent 250 & 69.56 & 25.92 & 15.51 \\
Norm-matched mean memories & 63.79 & 23.01 & 13.33 \\
Remove query & 11.04 & 4.30 & 2.79 \\
\bottomrule
\end{tabular*}
\end{table}

Table~\ref{tab:diagnostics} changes inputs at a fixed checkpoint without retraining. Cross-user history swap replaces the history with another user's sequence; order interventions retain the same multiset of items. Recent-history interventions retain the indicated suffix. The norm-matched control replaces item-specific memory directions with the catalog mean while preserving the original slot norms. Removing the query tests the fitted model's reliance on the current query.

We compute paired confidence intervals with 2,000 user-cluster bootstrap draws over 2,984 users and 2,999 requests. For cross-user history swap, the original-history Recall@100 advantage is 3.40 percentage points, with a 95\% interval of [2.37, 4.39]; removing history yields a 3.70-point advantage, with [2.69, 4.70]. For norm-matched mean memories, the advantage is 5.90 points, with [4.84, 6.96]. Across the three within-history permutations, NDCG@100 declines by 0.35--0.52 points and MRR@100 by 0.36--0.51 points; their paired intervals exclude zero. Recall@100 decreases by 0.13--0.67 points.

\subsection{History-Length and Item-Frequency Groups}
\label{app:subgroups}
\begin{table}[!htbp]
\centering
\caption{Recall@100 (\%) by retained history length and training frequency on the same 2,999 requests. Dense denotes Qwen3-Embedding-4B. Best values per row are bold.}
\label{tab:subgroups}
\small\setlength{\tabcolsep}{5pt}
\begin{tabular*}{\columnwidth}{@{\extracolsep{\fill}}lrrr@{}}
\toprule
Group & Requests & CRIM & Dense \\
\midrule
History 1--50 & 116 & \textbf{81.03} & 70.69 \\
History 51--100 & 450 & \textbf{78.22} & 65.78 \\
History 101--250 & 1,540 & \textbf{76.04} & 68.70 \\
History 251--500 & 893 & 52.97 & \textbf{67.53} \\
Frequency 1--10 & 7 & \textbf{100.00} & 57.14 \\
Frequency 11--100 & 1,718 & \textbf{72.47} & 68.86 \\
Frequency $>100$ & 1,274 & 65.78 & \textbf{66.88} \\
\bottomrule\end{tabular*}\end{table}

Table~\ref{tab:subgroups} groups the shared requests by retained history length and target frequency in training histories plus training targets. CRIM has higher recall in the groups through 250 items, whereas the dense reference leads in the 251--500 group. The aggregate paired Recall@100 difference is 1.70 points, with a 95\% interval of [$-0.60$, 4.03]. Only seven requests fall in the frequency 1--10 group.

\end{document}